\documentclass[twocolumn,english,superscriptaddress]{revtex4-1}
\usepackage[T1]{fontenc}
\usepackage[latin9]{inputenc}
\usepackage{color}
\usepackage{textcomp}
\usepackage{amstext}
\usepackage{graphicx}
\usepackage{subscript}

\makeatletter

\newcommand{\lyxmathsym}[1]{\ifmmode\begingroup\def\b@ld{bold}
  \text{\ifx\math@version\b@ld\bfseries\fi#1}\endgroup\else#1\fi}

\usepackage{hyperref}
\usepackage{upgreek,textgreek}
\hypersetup{colorlinks=true,linkcolor=blue,citecolor=blue,urlcolor=blue}
\makeatother

\usepackage{babel}
\begin{document}
\title{Stacking faults in $\alpha$-RuCl\textsubscript{3} revealed by local electric polarization}
\author{Xinrun Mi}
\affiliation{Low Temperature Physics Lab, College of Physics \& Center of Quantum
Materials and Devices, Chongqing University,Chongqing 401331, China}
\author{Xiao Wang}
\thanks{Xinrun Mi and Xiao Wang contributed equally to this work.}
\affiliation{J\"ulich Centre for Neutron Science (JCNS) at Heinz Maier-Leibnitz
Zentrum (MLZ), Forschungszentrum J\"ulich GmbH, Lichtenbergstr. 1,
D-85747 Garching, Germany}
\author{Hengrui Gui}
\affiliation{Low Temperature Physics Lab, College of Physics \& Center of Quantum
Materials and Devices, Chongqing University,Chongqing 401331, China}
\author{Maochai Pi}
\affiliation{Low Temperature Physics Lab, College of Physics \& Center of Quantum
Materials and Devices, Chongqing University,Chongqing 401331, China}
\author{Tingting Zheng}
\affiliation{Low Temperature Physics Lab, College of Physics \& Center of Quantum
Materials and Devices, Chongqing University,Chongqing 401331, China}
\author{Kunya Yang}
\affiliation{Low Temperature Physics Lab, College of Physics \& Center of Quantum
Materials and Devices, Chongqing University,Chongqing 401331, China}
\author{Yuhan Gan}
\affiliation{Low Temperature Physics Lab, College of Physics \& Center of Quantum
Materials and Devices, Chongqing University,Chongqing 401331, China}
\author{Peipei Wang}
\affiliation{Department of Physics, Southern University of Science and Technology,
518055 Shenzhen, China}
\author{Alei Li}
\affiliation{Department of Physics, Southern University of Science and Technology,
518055 Shenzhen, China}
\author{Aifeng Wang}
\affiliation{Low Temperature Physics Lab, College of Physics \& Center of Quantum
Materials and Devices, Chongqing University,Chongqing 401331, China}
\author{Liyuan Zhang}
\affiliation{Department of Physics, Southern University of Science and Technology,
518055 Shenzhen, China}
\author{Yixi Su}
\affiliation{J\"ulich Centre for Neutron Science (JCNS) at Heinz Maier-Leibnitz
Zentrum (MLZ), Forschungszentrum J\"ulich GmbH, Lichtenbergstr. 1,
D-85747 Garching, Germany}
\author{Yisheng Chai}
\affiliation{Low Temperature Physics Lab, College of Physics \& Center of Quantum
Materials and Devices, Chongqing University,Chongqing 401331, China}
\author{Mingquan He}
\email{mingquan.he@cqu.edu.cn}

\affiliation{Low Temperature Physics Lab, College of Physics \& Center of Quantum
Materials and Devices, Chongqing University,Chongqing 401331, China}
\date{04/28/21}
\begin{abstract}
We present out-of-plane dielectric and magnetodielectric measurements
of single crystallines $\alpha$-RuCl\textsubscript{3} with various
degrees of stack faults. A frequency dependent, but
field independent, dielectric anomaly appears at $T_{A}\:(f=100\:\mathrm{kHz})\sim$
4 K once both magnetic transitions at $T_{N1}\sim$ 7 K and $T_{N2}\sim$
14 K set in. The observed dielectric anomaly is attributed
to the emergency of possible local electric polarizations whose inversion
symmetry is broken by inhomogeneously distributed stacking faults.
A field-induced intermediate phase is only observed when a magnetic
field is applied perpendicular to the Ru-Ru bonds for samples with
minimal stacking faults. Less pronounced in-plane anisotropy is found
in samples with sizable contribution from stacking imperfections.
Our findings suggest that dielectric measurement is a  sensitive
probe in detecting the structural and magnetic properties, which may
be a promising tool especially in studying $\alpha$-RuCl\textsubscript{3} thin film devices. Moreover, the stacking
details of RuCl\textsubscript{3} layers strongly affect the ground
state both in the magnetic and electric channels. Such a fragile ground
state against stacking faults needs to be overcome for realistic applications
utilizing the magnetic and/or electric properties of Kitaev based
physics in $\alpha$-RuCl\textsubscript{3}.
\end{abstract}
\maketitle

\section{Introduction}

The possible emergence of Majorana fermion excitations in a $S=1/2$ Kitaev
quantum spin liquid (QSL) state is thought to be a promising channel
to realize topological quantum computing \citep{KITAEV20062,Nayak2008}.
Solvability of the honeycomb Kitaev model has boosted the experimental
interests to search for real materials in which Kitaev physics is at
play. Some Mott insulators with strong spin-orbit coupling induced
effective spin-1/2 ($J_{eff}=1/2$) appear to be promising candidates
of Kitaev materials \citep{Broholm:2020aa_review,Trebst2017_review,Wen:2019aa}.
To date, a few possible compounds including iridates ($A$\textsubscript{2}IrO\textsubscript{3}, $A$=Li, Na) \citep{Singh2012_A2IrO3,Modic:2014aa,Kimchi2014_Li2IrO3,Takayma2015_Li2IrO3,Glamazda:2016aa,Sigh2010}
and ruthenates ($\alpha$-RuCl\textsubscript{3}) \citep{Banerjee:2016NatM,Cao2016_stacking,Glamazda2017_Raman_hysteresis,Johnson2015_zigzag,Kubota2015_sus_hysteresis,Park2016_x_ray_hysteresis,Wang2017_pressure,Baek2017_NMR,Banerjee2017,Majumder2015,Kasahara:2018aa}
have been found, although conclusive evidence is still lacking.

Of particular interest is the layered compound $\alpha$-RuCl\textsubscript{3},
which is in close proximity to the ideal Kitaev model although a zigzag
antiferromagnetic (AF) order is favored at low temperatures due to
sizable Heisenberg interactions and non-zero off diagonal terms \citep{Banerjee:2016NatM,Norman2106_review,Zhou2017_review,Trebst2017_review}.
The AF ground state is unstable against application of pressure \citep{Wang2017_pressure,Cui2017_Press_NMR,He_2018}
and in-plane magnetic fields \citep{Baek2017_NMR,Banerjee2017_field,Gass2020,Balz2019,Lampen-Kelley2018,Ran2017}.
The system enters a quantum disordered phase when an in-plane magnetic
field higher than a critical value of $\mu_{0}H_{c}\sim$ 8 T is applied.
Evidence of fractional excitations emerged from this field-induced
quantum disordered state has been reported by various techniques \citep{Banerjee2017_field,Kasahara:2018aa,Do:2017aaHeatcapacity},
but the nature of this phase is still under debate. Meanwhile, the
magnetic transition is very sensitive to the stacking sequence of the honeycomb
layers. It has been suggested that a stacking of $ABC$
series produces the AF transition at $T_{N1}\sim$ 7 K, and that an $ABAB$
stacking is responsible for the transition at $T_{N2}\sim$ 14 K
\citep{Cao2016_stacking}. Neutron scattering experiments indeed found
non-negligible interlayer magnetic interactions \citep{Balz2019,Balz2020},
which necessarily couple to the details of layer stacking. Stacking faults
can be easily formed due to weak interlayer van-der-Waals bonding
(< 1 meV \citep{Kim2016_weak_bonding}) and the small energy difference
between these two configurations. Formation of stacking faults has
been evidenced by the appearance of multiple magnetic transitions within
one sample \citep{Cao2016_stacking,Kubota2015_sus_hysteresis}. To
date, big progress has been made to unravel the interplay between
magnetic, lattice and Kitaev interactions both theoretically and experimentally.
However, less effort has been applied to the study of stacking faults and their
related magnetic properties from the charge and/or electric degrees
of freedom, which are also important aspects to understand the ground
state and for future real device applications.

In this article, we report on the out-of-plane dielectric and magnetodielectric
measurements of $\alpha$-RuCl\textsubscript{3} single crystals with
different degrees of stacking faults. A frequency
dependent dielectric anomaly is observed at $T_{A}\:(f=100\:\mathrm{kHz})\sim$
4 K in samples showing both magnetic transitions at $T_{N1}\sim$
7 K and $T_{N2}\sim$ 14 K. Suppression of the magnetic transitions
using in-plane magnetic fields produces negligible effect on the observed
dielectric anomaly. No signature of such a dielectric anomaly is found
in crystals with the dominant transition at $T_{N1}\sim$ 7 K. We conclude
that the observed dielectric anomaly likely originates from local
electric polarizations, which appear at the interfaces between $ABC$ and
$ABAB$ stacking. The inversion symmetry is possibly broken by inhomogeneous
distribution of stacking faults. The magnetodielectric effect is
found to be anisotropic for in-plane magnetic fields applied parallel
and perpendicular to the Ru-Ru bonds. The field-induced intermediate
state found in earlier reports with magnetic fields applied perpendicular
to the Ru-Ru bonds \citep{Balz2019,Balz2020}, is confirmed by our
magnetodielectric measurements for samples with minimal stacking faults.
Our results suggest that the complex magnetic phases in $\alpha$-RuCl\textsubscript{3}
can also be accessed by dielectric probe, and that the structural
details determine the ground state both in the magnetic and electric
channels.

\begin{figure*}
\begin{centering}
\includegraphics[scale=0.6]{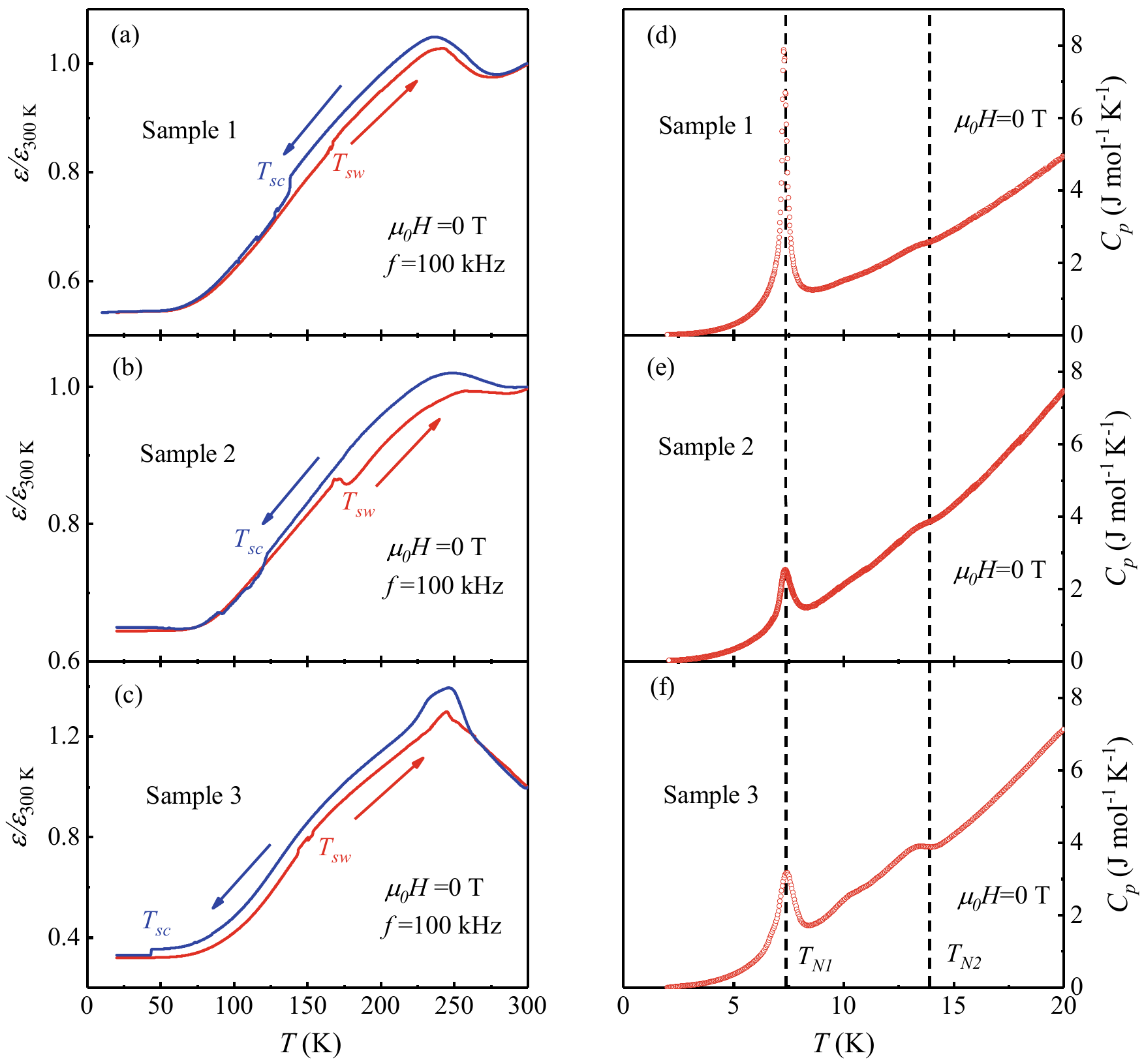}
\par\end{centering}
\caption{(a-c) Temperature dependence of the normalized dielectric constant, $\varepsilon(T)/\varepsilon_{300\mathrm{\,K}}$,
measured at zero field for three typical samples. The electric field
was applied perpendicular to the $ab$ plane ($\mathbf{E}\:||\:c\perp ab$
in the $R\bar{3}$ representation). A sample dependent hysteretic
structural transition is observed in all three samples using a temperature
ramping rate of 1 K/min ($T_{sc}$: transition temperature for cooling,
$T_{sw}$: transition temperature upon warming). (d-f) Temperature
evolution of specific heat near the magnetic transitions for the corresponding three samples. Vertical \textit{dash lines }in (d-f) mark the magnetic transitions.\label{fig:1}}
\end{figure*}

\section{Methods}

$\alpha$-RuCl\textsubscript{3} single crystals were grown by chemical vapour transport method in a two-zone furnace. Commercial RuCl\textsubscript{3} powders (3 g in mass, Furuya metal) were first sealed in a quartz tube (length: 12 cm, diameter: 2 cm) and then put in the two-zone furnace (source temperature 790 $\lyxmathsym{\textdegree}$C,
sink temperature 710 $\lyxmathsym{\textdegree}$C). Black shiny plates of $\alpha$-RuCl\textsubscript{3} single crystals would appear at the sink end after dwelling for 5 days. The samples used in this study have typical dimension of
5\texttimes 5\texttimes 0.3 mm\textsuperscript{3}. The dielectric
constant of the $\alpha$-RuCl\textsubscript{3} single crystals was
measured by an Agilent E4980A LCR meter with electric field applied
perpendicular to the $ab$ plane ($\mathbf{E}\parallel c$ in the
$R\bar{3}$ notation) in a 9 T Quantum Design Dynacool System and
a 14 T Oxford Cryostat. The electrodes were prepared by sputtering
50 nm Au on both sides of the $ab$ crystal surfaces. The heat capacity
measurements were performed in a 9 T Quantum Design Dynacool System.

\section{Results And Discussions}

\subsection{Structural and Magnetic Transitions}

The normalized dielectric constants, $\varepsilon(T)/\varepsilon_{300\mathrm{\,K}}$,
of three typical $\alpha$-RuCl\textsubscript{3} single crystals
measured at zero magnetic field are presented in Figure \ref{fig:1}(a-c).
The electric field was applied perpendicular to the
$ab$ plane ($\mathbf{E}\parallel c\perp ab$). The room temperature
relative dielectric constant, $\varepsilon_{r}(300\:\mathrm{K})=\varepsilon/\varepsilon_{0}=Cd/\varepsilon_{0}A$,
was estimated to be $\sim$15 for all three samples. Here, $\varepsilon_{0}$
is the dielectric constant of vaccum, $C$ is the measured capacitance,
$d$ is the sample thickness, $A$ is the effective area of the elelctrodes
deposited on sample surfaces. Hysteretic step-like
features appearing at $T_{sc}$ (upon cooling) and $T_{sw}$ (upon
warming) are clearly seen in all samples, which are
signatures of a first-order structural transition. Evidence of such
a structural transition has also been captured by other techniques,
including heat capacity \citep{Widmann2019_Thermalexpansion}, magnetization
\citep{Kubota2015_sus_hysteresis}, X-ray \citep{Park2016_x_ray_hysteresis}, Raman scattering \citep{Glamazda2017_Raman_hysteresis}, thermal-expansion
\citep{He_2018,Gass2020,Widmann2019_Thermalexpansion,Reschke_2018_thermalexpansion}
and an earlier dielectric study \citep{Zheng:2018aa}. Instead of 
the high-temperature monoclinic $C2/m$ structure, it is likely that
a different phase is formed at low temperatures (trigonal $P3_{1}12$
or rhombohedral $R\bar{3}$ phase \citep{Stroganov1957,Ran2017,Park2016_x_ray_hysteresis}).
However, the origin of this structural transition remains unclear.
Here we adopt the $R\bar{3}$ convention for convenience. Given the
nature of weak van-der-Waals bonding, formation of stacking faults
is inevitable during such a hysteretic transition. Only one dominant
magnetic transition occurring at either $T_{N1}\sim$
7 K or $T_{N2}\sim$ 14 K would be expected if a crystal has minimal
stacking faults. As shown in Fig. \ref{fig:1}(d), this case is realized
in Sample 1 which shows a dominant sharp peak at $T_{N1}=$ 7.3 K
in the specific heat data. For Sample 2 and Sample 3, another step-like
feature appears at $T_{N2}=$ 14 K in addition to
the major peak locating at $T_{N1}$ {[}see Figs.\ref{fig:1}(e) and
\ref{fig:1}(f){]}. This simply indicates a considerable mixing of
$ABC$ and $ABAB$ polymorphs and sizable stacking faults \citep{Cao2016_stacking}. 
Different degrees of stacking faults in the three samples studied here are likely caused by careless handling during electrodes preparation \citep{Cao2016_stacking}. Note that a broad hump sitting around 10 K is also visible in Sample
3. Similar feature has been reported earlier \citep{Majumder2015,He_2018,Kubota2015_sus_hysteresis},
whose origin is attributed to competing exchange interactions \citep{Kubota2015_sus_hysteresis}.
Although the exact nature is yet to be clarified, sizable stacking
faults certainly play an important role.

The width of the hysteretic structural transition $T_{s,width}=T_{sw}-T_{sc}$,
also correlates with the degree of stacking faults. Compared to Sample
1, more and more stacking faults are introduced into Sample 2 and
Sample 3 as evidenced by the heat capacity results. Correspondingly,
$T_{s,width}$ spans over a larger and larger temperature range for
Sample 2 and Sample 3. It is likely that when cooled below $T_{sc}$,
the transformation between the high-temperature and low-temperature
phases is less complete for Sample 2 and Sample 3 than for Sample
1 \citep{He_2018,Reschke_2018_thermalexpansion}. Naturally, more stacking faults are formed and a larger hysteresis is necessary to restore the high temperature structure
upon warming. Note that the dielectric constant is generally not fully
recovered after $T_{sw}$ upon warming, although the lattice constants
are more or less restored \citep{He_2018,Gass2020,Widmann2019_Thermalexpansion,Reschke_2018_thermalexpansion}.
This implies that the formed stacking faults are irreversible when
passing through the structural transition, and that the dielectric
probing is sensitive to these minute structural perturbations.
Reversible dielectric constant is eventually achieved near room temperature
when thermal energy is strong enough to wipe out these tiny structural
differences.

\subsection{Dielectric Anomaly}

\begin{figure*}
\begin{centering}
\includegraphics[scale=0.63]{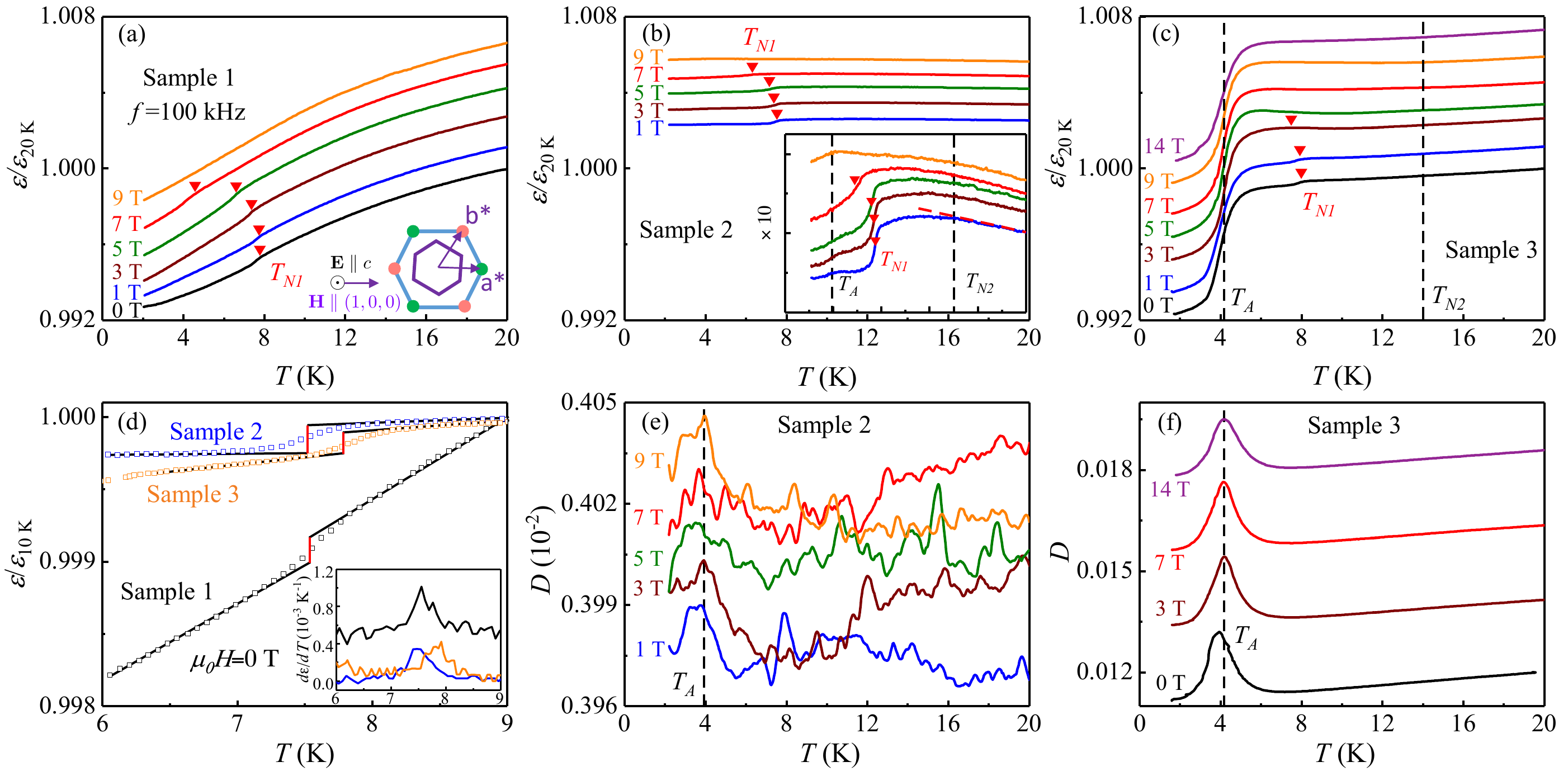}
\par\end{centering}
\caption{(a-c) Temperature evolution of, $\varepsilon/\varepsilon_{20\,\mathrm{K}}$,
recorded at $f=100$ kHz. In-plane magnetic fields were applied along
the reciprocal space $(1\text{,}0,0)$ direction {[}parallel to the
real space Ru-Ru bonds, see inset in (a){]}. Curves have been shifted
vertically for clarity. (d) The enlarged view of $\varepsilon/\varepsilon_{10\mathrm{\,K}}$
near $T_{N1}$. The empty squares are experimental raw data and the lines are guides to the eyes. The inset of (d) shows $d\varepsilon/dT$. (e) and (f) The
dissipation (D) as a function of temperature measured in fixed fields
of Sample 2 and Sample 3, respectively. Inset in (a): an illustration
of the in-plane reciprocal space (inner purple hexagon) and real space
(outer blue hexagon) representations. Solid triangles in (a-c) track
the variation of $T_{N1}$ with respect to in-plane magnetic fields.
Vertical \textit{dash lines} in (b) and (c) mark the dielectric anomaly
and the transition at $T_{N2}$ {[}determined by specific heat data
in Fig. \ref{fig:1}(e,f){]}.\label{fig:2}}
\end{figure*}

\begin{figure*}
\begin{centering}
\includegraphics[scale=0.68]{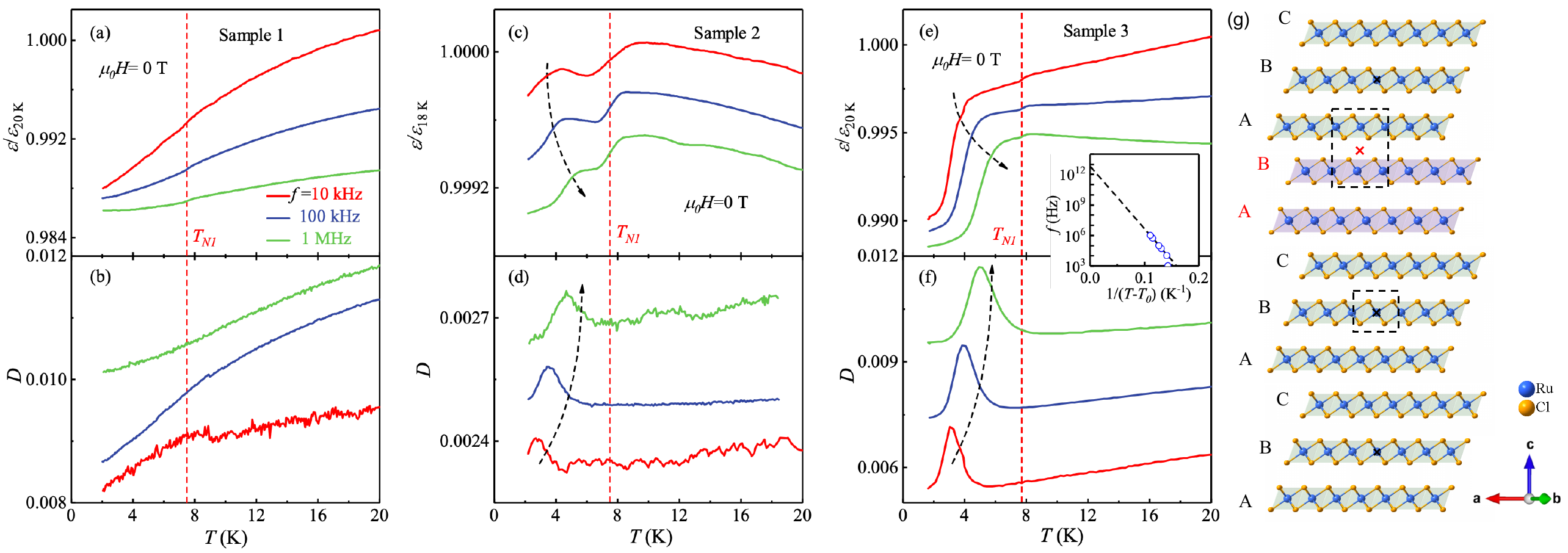}
\par\end{centering}
\caption{Frequency dependency of (a)(c)(e): $\varepsilon/\varepsilon_{20\,\mathrm{K}}$ and (b)(d)(f): the corresponding
dissipation (D) of all samples measured in zero field. The dielectric
anomaly depends strongly on frequency and $T_{A}$ shifts gradually
to higher temperatures for increasing frequency. (g) A schematic
illustration of inversion symmetry breaking caused by non-uniformly
distributed stacking faults. Black (red) cross marked out by dash
rectangle labels the inversion center of each $ABC$ ($AB$) unit.
The inversion counterpart is missing for the bottom $ABC$ segment
when an $AB$ unit is inserted asymmetrically. Inset in (e) and (f)
is a fitting of the frequency dependency of $T_{A}$ according to
the Vogel-Fulcher law $f=f_{0}exp\left[-E/k_{B}(T_{A}-T_{0})\right]$
using the Boltzmann constant $k_{B}$, a temperature constant $T_{0}=-4$ K, an activation energy $E=143$ K and a characteristic
frequency $f_{0}=7.6\times10^{12}$ Hz.\label{fig:3}}
\end{figure*}

In Fig. \ref{fig:2}, we present the temperature dependence of the
normalized dielectric constant, $\varepsilon/\varepsilon_{20\,\mathrm{K}}$,
near the AF transitions region. The magnetic fields were applied along
the Ru-Ru bonds {[}$\mathbf{H}\parallel(1\text{,}0,0)$ in the reciprocal
space, see inset of Fig. \ref{fig:2}(a){]}. No significant differences
are found when applying magnetic fields perpendicular to the Ru-Ru
bonds for temperature sweep (data not shown here). Clearly, these
three samples have different dielectric response with respect to temperatures
and external in-plane magnetic fields. $\varepsilon/\varepsilon_{20\,\mathrm{K}}$
of Sample 1 decreases continuously with cooling in all fields {[}Fig.
\ref{fig:2}(a){]}. At zero field, only a step-like jump appears at $T_{N1}=$
7.5 K {[}determined by the peak in $d\varepsilon/dT$, see the inset
of Fig. \ref{fig:2}(d){]}, below which the dielectric constant is
slightly reduced. This magnetic order induced dielectric reduction
is a possible signature of a type-II multiferroic as discussed by J.
Zheng \textit{et al}. \citep{Zheng:2018aa}. Another more trivial
explanation could be simple enhancement of the $c$-axis lattice constant
caused by magnetoelastic coupling as found by thermal-expansion measurements
\citep{Gass2020,He_2018}. Upon application of in-plane magnetic fields,
the AF transition at $T_{N1}$ shifts gradually towards lower temperatures
and eventually varnishes above the critical field $\mu_{0}H_{c1}\sim8$
T. Suppression of $T_{N1}$ by in-plane magnetic fields has also been
detected by other techniques, which is thought to be an prominent
way for driving the system into the QSL state \citep{Baek2017_NMR,Banerjee2017_field,Gass2020,Balz2019,Lampen-Kelley2018,Ran2017}.

In contrast to Sample 1, a negative linear slope is found in Sample
2 {[}see the inset of Fig. \ref{fig:2}(b){]} when cooled down from 20 K. This linear
temperature dependence deviates below $T_{N2}\sim$ 14 K, indicating
the appearance of the AF order in the $ABAB$ polymorph. Further cooling
leads to a step-like jump at $T_{N1}=$ 7.5 K in small fields. Note
that the size of the dielectric reduction occurring at $T_{N1}$ is similar
in all three samples, as shown in Fig. \ref{fig:2}(d), but the transition
appears to be sharpest in Sample 1 as expected {[}see the inset of
Fig. \ref{fig:2}(d){]}. In addition to the magnetic transitions,
a dielectric anomaly emerges at a lower temperature $T_{A}\sim4$ K
for $f=100$ kHz, which is absent in Sample 1. As illustrated in Fig.
\ref{fig:2}(e), this dielectric anomaly is accompanied by sizable
dissipation, and $T_{A}$ is defined as the temperature where the dissipation
shows a peak. The dielectric anomaly becomes less visible in the intermediate
field region (about 5 T to 7 T) as $T_{N1}$ is pushed gradually towards
$T_{A}$. However, a clear signature is recovered again at $\mu_{0}H=9$
T once the AF order formed below $T_{N1}$ is fully suppressed. The
dissipation peak {[}Fig. \ref{fig:2}(e){]}, however, stays nearly
intact throughout the studied field range. This implies little effect
on the dielectric anomaly by applying in-plane magnetic fields.

The dielectric anomaly becomes the dominant feature in Sample 3 {[}Fig.
\ref{fig:2}(c){]}, and a significant amount of dielectric constant is
lost at $T_{A}$. A clear peak is seen in the dissipation at $T_{A}$, as shown in Fig. \ref{fig:2}(f), which is field independent up to 14 T. As indicated by the specific heat data
(see Fig. \ref{fig:1}), Sample 2 and Sample 3 have a higher degree of stacking faults. Thus, the observed dielectric anomaly correlates
strongly with the degree of stacking faults. An earlier study performed
by T. Aoyama \textit{et al}. \citep{Takuya2017} reported such a dielectric
anomaly for both $\mathbf{E}\parallel c$ and $\mathbf{E}\parallel ab$
in samples with major transition at $T_{N2}\sim$ 14 K. On the other
hand, J. Zheng \textit{et al}. \citep{Zheng:2018aa} only found a
step-like dielectric reduction at $T_{N1}\sim$ 7 K for $\mathbf{E}\parallel c$
in samples with minimal stacking faults. T. Aoyama \textit{et al}.
\citep{Takuya2017} suggested that zigzag AF order induced local polarizations
are responsible for the observed dielectric anomaly in the case of
$\mathbf{E}\parallel ab$. However, unlike the AF transitions which
fade away above $\mu_{0}H_{c2}\sim10$ T {[}see \citep{Takuya2017,Kubota2015_sus_hysteresis}
and Fig. 4(e,f){]}, the observed dielectric anomaly for $\mathbf{E}\parallel c$
appears to be robust up to 14 T, as shown in Figs. \ref{fig:2}(c) and \ref{fig:2}(f).
Moreover, the dielectric loss for $\mathbf{E}\parallel c$ below $T_{A}$
is much more profound in Sample 3 (about $5\times10^{-3}$) than that
of crystals with less stacking faults (negligible dielectric loss
in Sample 1 and Sample 2, $\sim5\times10^{-4}$ dielectric loss in
samples with dominant $ABAB$ stacking \citep{Takuya2017}). Therefore,
it is likely that the dielectric anomaly found here for $\mathbf{E}\parallel c$
is more closely associated with stacking faults than the zigzag AF
order.

To further unravel the nature of the observed dielectric anomaly,
its frequency dependency is explored in
Figure \ref{fig:3}. Clearly, the dielectric anomaly is strongly frequency
dependent. As shown in Fig. \ref{fig:3}(c)-(f), $T_{A}$ moves monotonically
towards higher temperatures for increasing frequency. Therefore,
any long-range order can be ruled out, as no frequency dependency would
be expected. Similar frequency dependency of the dielectric anomaly is
also found for $\mathbf{E}\parallel ab$, and its origin is attributed
to a glassy state of zigzag AF order induced local electric polarizations
\citep{Takuya2017}. As displayed in the inset of Figs. \ref{fig:3}(c) and
\ref{fig:3}(f), the frequency dependency of $T_{A}$ can be well
described by the Vogel-Fulcher law $f=f_{0}exp\left[-E/k_{B}(T_{A}-T_{0})\right]$
with similar $T_{0}=-4$ K, activation energy $E=143$ K and characteristic
frequency $f_{0}=7.6\times10^{12}$ Hz compared to those of $\mathbf{E}\parallel ab$
\citep{Takuya2017}. This suggests that the same physics is at play
for both $\mathbf{E}\parallel ab$ and $\mathbf{E}\parallel c$, i.e.,
a possible glassy state of local electric polarizations is formed. Such an argument is plausible as the characteristics of dipolar glasses \citep{Courtens1984}, spin glasses \citep{BHOWMIK2002101} and relaxor ferroelectrics \citep{Bokov:2006aa} can be described by similar phenomenological Vogel-Fulcher approach.
Given the negligible response to in-plane
magnetic fields, the electric polarizations here for $\mathbf{E}\parallel c$
are most probably caused by stacking faults located at the $ABC/ABAB$
interfaces. As sketched in Fig. \ref{fig:3}(g), the inversion center
sits in the middle of $B$ layer in each $ABC$ unit of pure $ABC$
stacking. The bottom $ABC$ segment could not find its inversion counterpart
if an $AB$ unit was inserted asymmetrically into $ABC$ layers. We
note that a uniform distribution of stacking faults still preserves
the inversion symmetry. Instead of zigzag AF order induced symmetry
breaking in the $ab$ plane, here along the $c$-axis, the inversion
symmetry is broken locally by inhomogeneously distributed stacking
faults.

\subsection{Magnetodielectric Effect and Field induced Intermediate State}

\begin{figure*}[t]
\begin{centering}
\includegraphics[scale=0.8]{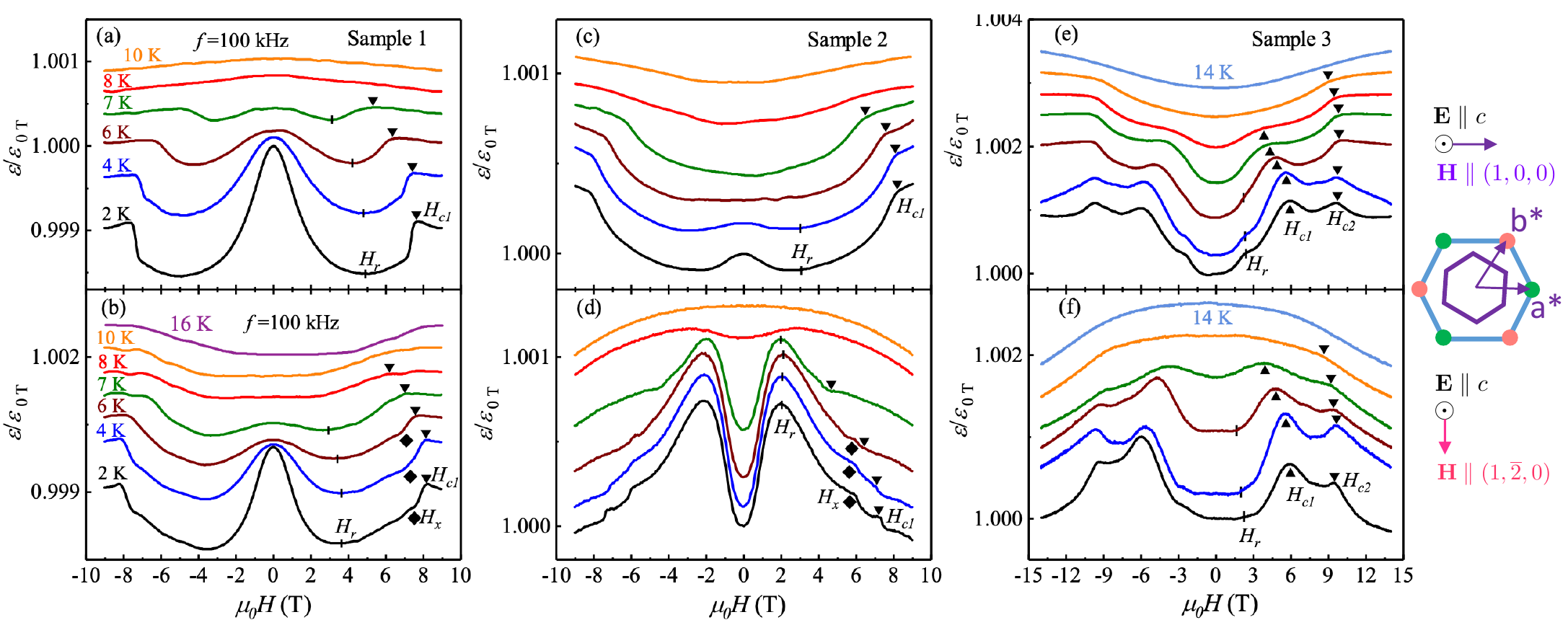}
\par\end{centering}
\caption{Magnetodielectric effect measured at fixed temperatures and constant
frequency $f$=100 kHz. In (a) and (b), magnetic fields were along and
perpendicular to the Ru-Ru bonds with $\mathbf{H}\parallel(1,0,0)$
and $\mathbf{H}\parallel(1,\overline{2},0)$, respectively for Sample
1. (c)(d) and (e)(f): same measurements with (a)(b) for Sample 2 and
Sample 3, respectively. Curves are shifted vertically for clarity.
\label{fig:4}}
\end{figure*}

\begin{figure*}[!t]
\begin{centering}
\includegraphics[scale=0.85]{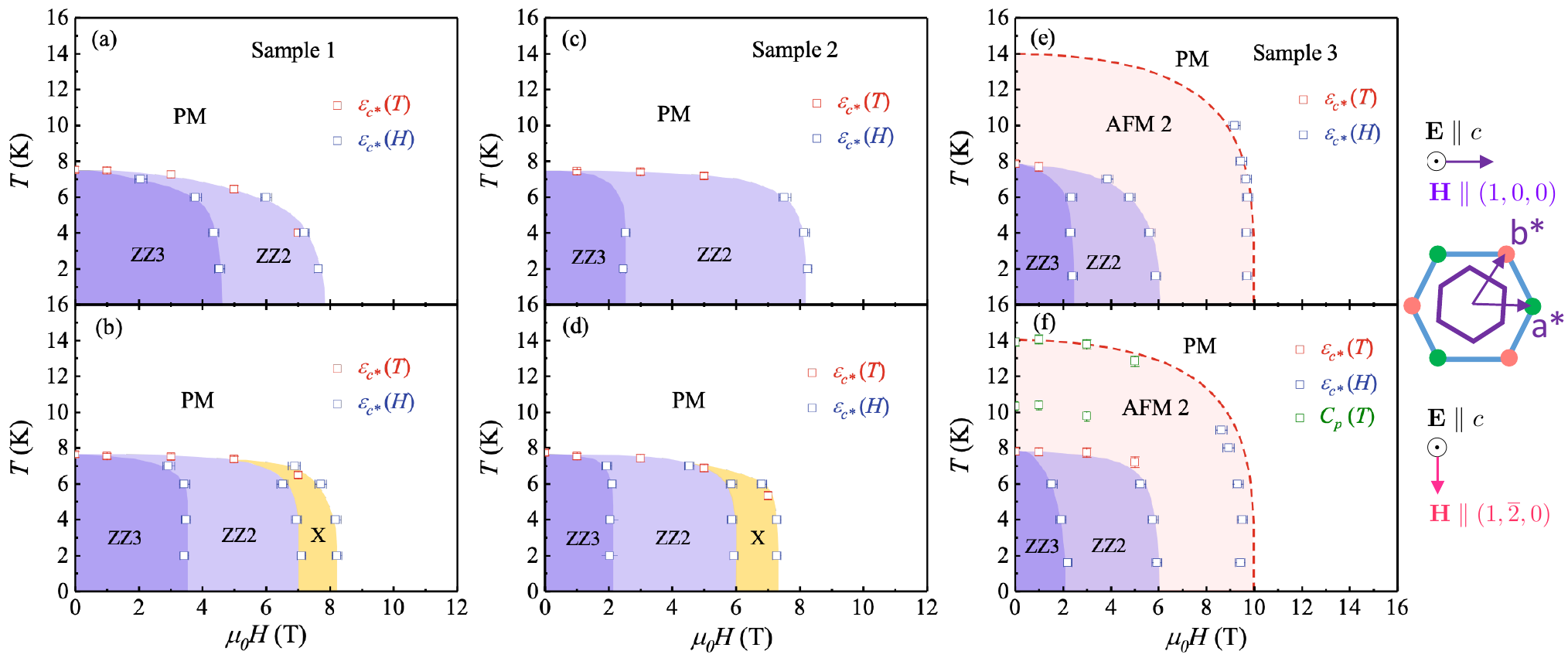}
\par\end{centering}
\caption{$T-H$ phase diagrams of the studied three samples derived from Fig.
\ref{fig:2} and Fig. \ref{fig:4}. (a) and (b): $T-H$ phase diagrams
of Sample 1 with fields applied along the $\mathbf{H}\parallel(1,0,0)$
and $\mathbf{H}\parallel(1,\overline{2},0)$, respectively. (c)(d)
and (e)(f): Same representations to those of (a)(b) for Sample 2 and
Sample 3, respectively. The symbols ZZ3 and ZZ2 represent three and
two domains in the zigzag phase, respectively. The field induced intermediate
state is labeled as the X phase. The paramagnetic state is labeled as PM and the zigzag AF order formed below $T_{N2}$ is represented as AFM 2. Green dots were obtained in specific
heat experiments.\label{fig:5}}
\end{figure*}

In this section, we study the anisotropic magnetodielectric effect
and the results are presented in Figure \ref{fig:4}. In-plane anisotropy
and multiple field induced magnetic phase transitions are clearly
seen in all samples. In the paramagnetic state above $T_{N1}$ or
$T_{N2}$, the magnetodielectric data generally shows temperature
independent backgrounds, which vary between samples and field directions.
This discrepancy might be caused by sample and thermal history dependent
stacking faults as measurements for different field orientations were
performed after several thermal cycles up to room temperature.

For Sample 1 shown in Figs. \ref{fig:4}(a) and \ref{fig:4}(b), a
local minimum at $\mu_{0}H_{r}\sim$ 4 T is observed consistently below $T_{N1}$
along both field directions. This dip feature is likely associated
with the domain repopulation as found by neutron scattering experiments \citep{Sears2017Domain}.
The domain reorientation apparently exists in all samples. The transition from the AF ordered state to the quantum disordered
phase is evidenced by a step-like jump sitting at $\mu_{0}H_{c1}\sim$ 8
T in Sample 1 and Sample 2. Additionally, a small bump is found at
$\mu_{0}H_{x}\sim$ 6 T for $\mathbf{H}\parallel(1,\overline{2},0)$ only
(perpendicular to the Ru-Ru bonds), which is more pronounced in Sample
2 {[}see Fig. \ref{fig:4}(d){]}. This field-induced intermediate
phase between $H_{x}$ and $H_{c1}$with field applied perpendicular
to the Ru-Ru bonds has also been evidenced by AC susceptibility and
inelastic neutron scattering experiments \citep{Balz2019,Balz2020},
although its origin still remains elusive.

Figures \ref{fig:4}(e) and \ref{fig:4}(f) display the results of
Sample 3. An additional peak appears at $\mu_{0}H_{c2}\sim$ 10 T
below $T_{N2}$, apart from familiar features at $\mu_{0}H_{r}$ and $\mu_{0}H_{c1}$.
Apparently, a higher field is necessary to partially align the zigzag
order emerged below $T_{N2}$ \citep{Kubota2015_sus_hysteresis,Takuya2017}.
Compared to other two samples, the critical field $\mu_{0}H_{c1}$ for suppressing
the AF zigzag order formed below $T_{N1}$ is reduced from $\sim$
8 T to $\sim$ 6 T at 2 K. The field induced intermediate phase is
not visible in Sample 3 due to the close proximity between $\mu_{0}H_{c1}$
and $\mu_{0}H_{x}$.

\subsection{Phase Diagram}

In Figure \ref{fig:5}, we summarize the $T-H$ phase diagrams of
the three different samples studied here. The data points were extracted
from the dielectric, magnetodielectric and specific heat results. One sees that all the
phases found by other techniques have been mapped out nicely using
dielectric probing. Similarities can be found in Sample 1 and Sample
2 where the magnetic phase is mainly governed by the AF zigzag order
formed below $T_{N1}$. Clear in-plane anisotropies are found for
these two samples. For $\mathbf{H}\parallel(1,0,0)$, main features
take place at the domain repopulation field $\mu_{0}H_{r}$ and the suppression
of AF order crossing the $\mu_{0}H_{c1}$ lines. An additional field induced
intermediate $X$ state shows up between $\mu_{0}H_{x}$ and $\mu_{0}H_{c1}$ for
$\mathbf{H}\parallel(1,\overline{2},0)$ in Sample 1 and Sample 2.
Less anisotropy appears in Sample 3 which has a sizable degree of stacking faults.
Both transitions at $\mu_{0}H_{r}$, $\mu_{0}H_{c1}$ and $\mu_{0}H_{c2}$ are identified,
whereas the intermediate state is missing due to the collapsing of
$\mu_{0}H_{c1}$ onto $\mu_{0}H_{x}$. The stacking faults induced glassy phase
of electric dipoles observed in Sample 2 and Sample 3 is not shown
in the phase diagram, as it is not a long-range order. By comparing
these three samples, it is clear that the ground state of $\alpha$-RuCl\textsubscript{3}
is very sensitive to structural details both in the magnetic and electric
channels due to strong interplay between lattice, spin and charge
degrees of freedom.

\section{Conclusions}

In summary, we observe an out-of-plane dielectric anomaly in $\alpha$-RuCl\textsubscript{3}
which appears to be closely associated with inhomogeneously distributed
stacking faults. This dielectric anomaly is a signature of a possible
glassy state of local electric polarizations, which is evidenced by
its strong frequency dependency and sizable dissipations. Immunity
to strong external in-plane magnetic fields of this glassy state points
to a structural origin instead of a magnetic root. Details of the magnetic
phase diagram, including the domain repopulation, the field-induced
intermediate (field applied perpendicular to the Ru-Ru bonds) and
the transition between AF ordered and disordered phases, were successfully
mapped out using dielectric and magnetodielectric probing. The dielectric
probing may thus serve as a promising tool for detecting stacking
faults and studying the magnetic properties of $\alpha$-RuCl\textsubscript{3}
thin film devices when bulk measurements are not accessible. Our findings
also suggest that the ground state is rather fragile against structural
perturbations both in the magnetic and electric channels. Further
efforts are needed to overcome this difficulty and to eventually manipulate
the novel Kitaev physics in $\alpha$-RuCl\textsubscript{3} based
devices.

\section*{Acknowledgments}

This work has been supported by National Natural Science Foundation
of China (Grant Nos.11904040, 12047564), Chongqing Research Program
of Basic Research and Frontier Technology, China (Grant No. cstc2020jcyj-msxmX0263),
Fundamental Research Funds for the Central Universities, China(2020CDJQY-A056,
2020CDJ-LHZZ-010, 2020CDJQY-Z006), Projects of President Foundation
of Chongqing University, China(2019CDXZWL002). Y. Chai acknowledges
the support by National Natural Science Foundation of China (Grant
No. 11674384, 11974065). A. Wang acknowledges the support by National
Natural Science Foundation of China (Grant No. 12004056).

\bibliographystyle{apsrev4-1}

\begin{thebibliography}{43}%
\makeatletter
\providecommand \@ifxundefined [1]{%
 \@ifx{#1\undefined}
}%
\providecommand \@ifnum [1]{%
 \ifnum #1\expandafter \@firstoftwo
 \else \expandafter \@secondoftwo
 \fi
}%
\providecommand \@ifx [1]{%
 \ifx #1\expandafter \@firstoftwo
 \else \expandafter \@secondoftwo
 \fi
}%
\providecommand \natexlab [1]{#1}%
\providecommand \enquote  [1]{``#1''}%
\providecommand \bibnamefont  [1]{#1}%
\providecommand \bibfnamefont [1]{#1}%
\providecommand \citenamefont [1]{#1}%
\providecommand \href@noop [0]{\@secondoftwo}%
\providecommand \href [0]{\begingroup \@sanitize@url \@href}%
\providecommand \@href[1]{\@@startlink{#1}\@@href}%
\providecommand \@@href[1]{\endgroup#1\@@endlink}%
\providecommand \@sanitize@url [0]{\catcode `\\12\catcode `\$12\catcode
  `\&12\catcode `\#12\catcode `\^12\catcode `\_12\catcode `\%12\relax}%
\providecommand \@@startlink[1]{}%
\providecommand \@@endlink[0]{}%
\providecommand \url  [0]{\begingroup\@sanitize@url \@url }%
\providecommand \@url [1]{\endgroup\@href {#1}{\urlprefix }}%
\providecommand \urlprefix  [0]{URL }%
\providecommand \Eprint [0]{\href }%
\providecommand \doibase [0]{http://dx.doi.org/}%
\providecommand \selectlanguage [0]{\@gobble}%
\providecommand \bibinfo  [0]{\@secondoftwo}%
\providecommand \bibfield  [0]{\@secondoftwo}%
\providecommand \translation [1]{[#1]}%
\providecommand \BibitemOpen [0]{}%
\providecommand \bibitemStop [0]{}%
\providecommand \bibitemNoStop [0]{.\EOS\space}%
\providecommand \EOS [0]{\spacefactor3000\relax}%
\providecommand \BibitemShut  [1]{\csname bibitem#1\endcsname}%
\let\auto@bib@innerbib\@empty
%</preamble>
\bibitem [{\citenamefont {Kitaev}(2006)}]{KITAEV20062}%
  \BibitemOpen
  \bibfield  {author} {\bibinfo {author} {\bibfnamefont {A.}~\bibnamefont
  {Kitaev}},\ }\href {\doibase http://dx.doi.org/10.1016/j.aop.2005.10.005}
  {\bibfield  {journal} {\bibinfo  {journal} {Ann. Phys.}\ }\textbf {\bibinfo
  {volume} {321}},\ \bibinfo {pages} {2 } (\bibinfo {year} {2006})}\BibitemShut
  {NoStop}%
\bibitem [{\citenamefont {Nayak}\ \emph {et~al.}(2008)\citenamefont {Nayak},
  \citenamefont {Simon}, \citenamefont {Stern}, \citenamefont {Freedman},\ and\
  \citenamefont {Das~Sarma}}]{Nayak2008}%
  \BibitemOpen
  \bibfield  {author} {\bibinfo {author} {\bibfnamefont {C.}~\bibnamefont
  {Nayak}}, \bibinfo {author} {\bibfnamefont {S.~H.}\ \bibnamefont {Simon}},
  \bibinfo {author} {\bibfnamefont {A.}~\bibnamefont {Stern}}, \bibinfo
  {author} {\bibfnamefont {M.}~\bibnamefont {Freedman}}, \ and\ \bibinfo
  {author} {\bibfnamefont {S.}~\bibnamefont {Das~Sarma}},\ }\href {\doibase
  10.1103/RevModPhys.80.1083} {\bibfield  {journal} {\bibinfo  {journal} {Rev.
  Mod. Phys.}\ }\textbf {\bibinfo {volume} {80}},\ \bibinfo {pages} {1083}
  (\bibinfo {year} {2008})}\BibitemShut {NoStop}%
\bibitem [{\citenamefont {Broholm}\ \emph {et~al.}(2020)\citenamefont
  {Broholm}, \citenamefont {Cava}, \citenamefont {Kivelson}, \citenamefont
  {Nocera}, \citenamefont {Norman},\ and\ \citenamefont
  {Senthil}}]{Broholm:2020aa_review}%
  \BibitemOpen
  \bibfield  {author} {\bibinfo {author} {\bibfnamefont {C.}~\bibnamefont
  {Broholm}}, \bibinfo {author} {\bibfnamefont {R.~J.}\ \bibnamefont {Cava}},
  \bibinfo {author} {\bibfnamefont {S.~A.}\ \bibnamefont {Kivelson}}, \bibinfo
  {author} {\bibfnamefont {D.~G.}\ \bibnamefont {Nocera}}, \bibinfo {author}
  {\bibfnamefont {M.~R.}\ \bibnamefont {Norman}}, \ and\ \bibinfo {author}
  {\bibfnamefont {T.}~\bibnamefont {Senthil}},\ }\href {\doibase
  10.1126/science.aay0668} {\bibfield  {journal} {\bibinfo  {journal}
  {Science}\ }\textbf {\bibinfo {volume} {367}},\ \bibinfo {pages} {eaay0668}
  (\bibinfo {year} {2020})}\BibitemShut {NoStop}%
\bibitem [{\citenamefont {Trebst}()}]{Trebst2017_review}%
  \BibitemOpen
  \bibfield  {author} {\bibinfo {author} {\bibfnamefont {S.}~\bibnamefont
  {Trebst}},\ }\href {http://arxiv.org/abs/1701.07056} {\ }\Eprint
  {http://arxiv.org/abs/1701.07056} {arXiv:1701.07056} \BibitemShut {NoStop}%
\bibitem [{\citenamefont {Wen}\ \emph {et~al.}(2019)\citenamefont {Wen},
  \citenamefont {Yu}, \citenamefont {Li}, \citenamefont {Yu},\ and\
  \citenamefont {Li}}]{Wen:2019aa}%
  \BibitemOpen
  \bibfield  {author} {\bibinfo {author} {\bibfnamefont {J.}~\bibnamefont
  {Wen}}, \bibinfo {author} {\bibfnamefont {S.-L.}\ \bibnamefont {Yu}},
  \bibinfo {author} {\bibfnamefont {S.}~\bibnamefont {Li}}, \bibinfo {author}
  {\bibfnamefont {W.}~\bibnamefont {Yu}}, \ and\ \bibinfo {author}
  {\bibfnamefont {J.-X.}\ \bibnamefont {Li}},\ }\href {\doibase
  10.1038/s41535-019-0151-6} {\bibfield  {journal} {\bibinfo  {journal} {npj
  Quantum Mater.}\ }\textbf {\bibinfo {volume} {4}},\ \bibinfo {pages} {12}
  (\bibinfo {year} {2019})}\BibitemShut {NoStop}%
\bibitem [{\citenamefont {Singh}\ \emph {et~al.}(2012)\citenamefont {Singh},
  \citenamefont {Manni}, \citenamefont {Reuther}, \citenamefont {Berlijn},
  \citenamefont {Thomale}, \citenamefont {Ku}, \citenamefont {Trebst},\ and\
  \citenamefont {Gegenwart}}]{Singh2012_A2IrO3}%
  \BibitemOpen
  \bibfield  {author} {\bibinfo {author} {\bibfnamefont {Y.}~\bibnamefont
  {Singh}}, \bibinfo {author} {\bibfnamefont {S.}~\bibnamefont {Manni}},
  \bibinfo {author} {\bibfnamefont {J.}~\bibnamefont {Reuther}}, \bibinfo
  {author} {\bibfnamefont {T.}~\bibnamefont {Berlijn}}, \bibinfo {author}
  {\bibfnamefont {R.}~\bibnamefont {Thomale}}, \bibinfo {author} {\bibfnamefont
  {W.}~\bibnamefont {Ku}}, \bibinfo {author} {\bibfnamefont {S.}~\bibnamefont
  {Trebst}}, \ and\ \bibinfo {author} {\bibfnamefont {P.}~\bibnamefont
  {Gegenwart}},\ }\href {\doibase 10.1103/PhysRevLett.108.127203} {\bibfield
  {journal} {\bibinfo  {journal} {Phys. Rev. Lett.}\ }\textbf {\bibinfo
  {volume} {108}},\ \bibinfo {pages} {127203} (\bibinfo {year}
  {2012})}\BibitemShut {NoStop}%
\bibitem [{\citenamefont {Modic}\ \emph {et~al.}(2014)\citenamefont {Modic},
  \citenamefont {Smidt}, \citenamefont {Kimchi}, \citenamefont {Breznay},
  \citenamefont {Biffin}, \citenamefont {Choi}, \citenamefont {Johnson},
  \citenamefont {Coldea}, \citenamefont {Watkins-Curry}, \citenamefont
  {McCandless}, \citenamefont {Chan}, \citenamefont {Gandara}, \citenamefont
  {Islam}, \citenamefont {Vishwanath}, \citenamefont {Shekhter}, \citenamefont
  {McDonald},\ and\ \citenamefont {Analytis}}]{Modic:2014aa}%
  \BibitemOpen
  \bibfield  {author} {\bibinfo {author} {\bibfnamefont {K.~A.}\ \bibnamefont
  {Modic}}, \bibinfo {author} {\bibfnamefont {T.~E.}\ \bibnamefont {Smidt}},
  \bibinfo {author} {\bibfnamefont {I.}~\bibnamefont {Kimchi}}, \bibinfo
  {author} {\bibfnamefont {N.~P.}\ \bibnamefont {Breznay}}, \bibinfo {author}
  {\bibfnamefont {A.}~\bibnamefont {Biffin}}, \bibinfo {author} {\bibfnamefont
  {S.}~\bibnamefont {Choi}}, \bibinfo {author} {\bibfnamefont {R.~D.}\
  \bibnamefont {Johnson}}, \bibinfo {author} {\bibfnamefont {R.}~\bibnamefont
  {Coldea}}, \bibinfo {author} {\bibfnamefont {P.}~\bibnamefont
  {Watkins-Curry}}, \bibinfo {author} {\bibfnamefont {G.~T.}\ \bibnamefont
  {McCandless}}, \bibinfo {author} {\bibfnamefont {J.~Y.}\ \bibnamefont
  {Chan}}, \bibinfo {author} {\bibfnamefont {F.}~\bibnamefont {Gandara}},
  \bibinfo {author} {\bibfnamefont {Z.}~\bibnamefont {Islam}}, \bibinfo
  {author} {\bibfnamefont {A.}~\bibnamefont {Vishwanath}}, \bibinfo {author}
  {\bibfnamefont {A.}~\bibnamefont {Shekhter}}, \bibinfo {author}
  {\bibfnamefont {R.~D.}\ \bibnamefont {McDonald}}, \ and\ \bibinfo {author}
  {\bibfnamefont {J.~G.}\ \bibnamefont {Analytis}},\ }\href
  {http://dx.doi.org/10.1038/ncomms5203} {\bibfield  {journal} {\bibinfo
  {journal} {Nat. Commun.}\ }\textbf {\bibinfo {volume} {5}},\ \bibinfo {pages}
  {4203} (\bibinfo {year} {2014})}\BibitemShut {NoStop}%
\bibitem [{\citenamefont {Kimchi}\ \emph {et~al.}(2014)\citenamefont {Kimchi},
  \citenamefont {Analytis},\ and\ \citenamefont
  {Vishwanath}}]{Kimchi2014_Li2IrO3}%
  \BibitemOpen
  \bibfield  {author} {\bibinfo {author} {\bibfnamefont {I.}~\bibnamefont
  {Kimchi}}, \bibinfo {author} {\bibfnamefont {J.~G.}\ \bibnamefont
  {Analytis}}, \ and\ \bibinfo {author} {\bibfnamefont {A.}~\bibnamefont
  {Vishwanath}},\ }\href {\doibase 10.1103/PhysRevB.90.205126} {\bibfield
  {journal} {\bibinfo  {journal} {Phys. Rev. B}\ }\textbf {\bibinfo {volume}
  {90}},\ \bibinfo {pages} {205126} (\bibinfo {year} {2014})}\BibitemShut
  {NoStop}%
\bibitem [{\citenamefont {Takayama}\ \emph {et~al.}(2015)\citenamefont
  {Takayama}, \citenamefont {Kato}, \citenamefont {Dinnebier}, \citenamefont
  {Nuss}, \citenamefont {Kono}, \citenamefont {Veiga}, \citenamefont {Fabbris},
  \citenamefont {Haskel},\ and\ \citenamefont {Takagi}}]{Takayma2015_Li2IrO3}%
  \BibitemOpen
  \bibfield  {author} {\bibinfo {author} {\bibfnamefont {T.}~\bibnamefont
  {Takayama}}, \bibinfo {author} {\bibfnamefont {A.}~\bibnamefont {Kato}},
  \bibinfo {author} {\bibfnamefont {R.}~\bibnamefont {Dinnebier}}, \bibinfo
  {author} {\bibfnamefont {J.}~\bibnamefont {Nuss}}, \bibinfo {author}
  {\bibfnamefont {H.}~\bibnamefont {Kono}}, \bibinfo {author} {\bibfnamefont
  {L.~S.~I.}\ \bibnamefont {Veiga}}, \bibinfo {author} {\bibfnamefont
  {G.}~\bibnamefont {Fabbris}}, \bibinfo {author} {\bibfnamefont
  {D.}~\bibnamefont {Haskel}}, \ and\ \bibinfo {author} {\bibfnamefont
  {H.}~\bibnamefont {Takagi}},\ }\href {\doibase
  10.1103/PhysRevLett.114.077202} {\bibfield  {journal} {\bibinfo  {journal}
  {Phys. Rev. Lett.}\ }\textbf {\bibinfo {volume} {114}},\ \bibinfo {pages}
  {077202} (\bibinfo {year} {2015})}\BibitemShut {NoStop}%
\bibitem [{\citenamefont {Glamazda}\ \emph {et~al.}(2016)\citenamefont
  {Glamazda}, \citenamefont {Lemmens}, \citenamefont {Do}, \citenamefont
  {Choi},\ and\ \citenamefont {Choi}}]{Glamazda:2016aa}%
  \BibitemOpen
  \bibfield  {author} {\bibinfo {author} {\bibfnamefont {A.}~\bibnamefont
  {Glamazda}}, \bibinfo {author} {\bibfnamefont {P.}~\bibnamefont {Lemmens}},
  \bibinfo {author} {\bibfnamefont {S.~H.}\ \bibnamefont {Do}}, \bibinfo
  {author} {\bibfnamefont {Y.~S.}\ \bibnamefont {Choi}}, \ and\ \bibinfo
  {author} {\bibfnamefont {K.~Y.}\ \bibnamefont {Choi}},\ }\href
  {http://dx.doi.org/10.1038/ncomms12286} {\bibfield  {journal} {\bibinfo
  {journal} {Nat. Commun.}\ }\textbf {\bibinfo {volume} {7}},\ \bibinfo {pages}
  {12286} (\bibinfo {year} {2016})}\BibitemShut {NoStop}%
\bibitem [{\citenamefont {Singh}\ and\ \citenamefont
  {Gegenwart}(2010)}]{Sigh2010}%
  \BibitemOpen
  \bibfield  {author} {\bibinfo {author} {\bibfnamefont {Y.}~\bibnamefont
  {Singh}}\ and\ \bibinfo {author} {\bibfnamefont {P.}~\bibnamefont
  {Gegenwart}},\ }\href {\doibase 10.1103/PhysRevB.82.064412} {\bibfield
  {journal} {\bibinfo  {journal} {Phys. Rev. B}\ }\textbf {\bibinfo {volume}
  {82}},\ \bibinfo {pages} {064412} (\bibinfo {year} {2010})}\BibitemShut
  {NoStop}%
\bibitem [{\citenamefont {Banerjee}\ \emph {et~al.}(2016)\citenamefont
  {Banerjee}, \citenamefont {Bridges}, \citenamefont {Yan}, \citenamefont
  {Aczel}, \citenamefont {Li}, \citenamefont {Stone}, \citenamefont {Granroth},
  \citenamefont {Lumsden}, \citenamefont {Yiu}, \citenamefont {Knolle},
  \citenamefont {Bhattacharjee}, \citenamefont {Kovrizhin}, \citenamefont
  {Moessner}, \citenamefont {Tennant}, \citenamefont {Mandrus},\ and\
  \citenamefont {Nagler}}]{Banerjee:2016NatM}%
  \BibitemOpen
  \bibfield  {author} {\bibinfo {author} {\bibfnamefont {A.}~\bibnamefont
  {Banerjee}}, \bibinfo {author} {\bibfnamefont {C.~A.}\ \bibnamefont
  {Bridges}}, \bibinfo {author} {\bibfnamefont {J.~Q.}\ \bibnamefont {Yan}},
  \bibinfo {author} {\bibfnamefont {A.~A.}\ \bibnamefont {Aczel}}, \bibinfo
  {author} {\bibfnamefont {L.}~\bibnamefont {Li}}, \bibinfo {author}
  {\bibfnamefont {M.~B.}\ \bibnamefont {Stone}}, \bibinfo {author}
  {\bibfnamefont {G.~E.}\ \bibnamefont {Granroth}}, \bibinfo {author}
  {\bibfnamefont {M.~D.}\ \bibnamefont {Lumsden}}, \bibinfo {author}
  {\bibfnamefont {Y.}~\bibnamefont {Yiu}}, \bibinfo {author} {\bibfnamefont
  {J.}~\bibnamefont {Knolle}}, \bibinfo {author} {\bibfnamefont
  {S.}~\bibnamefont {Bhattacharjee}}, \bibinfo {author} {\bibfnamefont {D.~L.}\
  \bibnamefont {Kovrizhin}}, \bibinfo {author} {\bibfnamefont {R.}~\bibnamefont
  {Moessner}}, \bibinfo {author} {\bibfnamefont {D.~A.}\ \bibnamefont
  {Tennant}}, \bibinfo {author} {\bibfnamefont {D.~G.}\ \bibnamefont
  {Mandrus}}, \ and\ \bibinfo {author} {\bibfnamefont {S.~E.}\ \bibnamefont
  {Nagler}},\ }\href {http://dx.doi.org/10.1038/nmat4604} {\bibfield  {journal}
  {\bibinfo  {journal} {Nat. Mater.}\ }\textbf {\bibinfo {volume} {15}},\
  \bibinfo {pages} {733} (\bibinfo {year} {2016})}\BibitemShut {NoStop}%
\bibitem [{\citenamefont {Cao}\ \emph {et~al.}(2016)\citenamefont {Cao},
  \citenamefont {Banerjee}, \citenamefont {Yan}, \citenamefont {Bridges},
  \citenamefont {Lumsden}, \citenamefont {Mandrus}, \citenamefont {Tennant},
  \citenamefont {Chakoumakos},\ and\ \citenamefont
  {Nagler}}]{Cao2016_stacking}%
  \BibitemOpen
  \bibfield  {author} {\bibinfo {author} {\bibfnamefont {H.~B.}\ \bibnamefont
  {Cao}}, \bibinfo {author} {\bibfnamefont {A.}~\bibnamefont {Banerjee}},
  \bibinfo {author} {\bibfnamefont {J.-Q.}\ \bibnamefont {Yan}}, \bibinfo
  {author} {\bibfnamefont {C.~A.}\ \bibnamefont {Bridges}}, \bibinfo {author}
  {\bibfnamefont {M.~D.}\ \bibnamefont {Lumsden}}, \bibinfo {author}
  {\bibfnamefont {D.~G.}\ \bibnamefont {Mandrus}}, \bibinfo {author}
  {\bibfnamefont {D.~A.}\ \bibnamefont {Tennant}}, \bibinfo {author}
  {\bibfnamefont {B.~C.}\ \bibnamefont {Chakoumakos}}, \ and\ \bibinfo {author}
  {\bibfnamefont {S.~E.}\ \bibnamefont {Nagler}},\ }\href {\doibase
  10.1103/PhysRevB.93.134423} {\bibfield  {journal} {\bibinfo  {journal} {Phys.
  Rev. B}\ }\textbf {\bibinfo {volume} {93}},\ \bibinfo {pages} {134423}
  (\bibinfo {year} {2016})}\BibitemShut {NoStop}%
\bibitem [{\citenamefont {Glamazda}\ \emph {et~al.}(2017)\citenamefont
  {Glamazda}, \citenamefont {Lemmens}, \citenamefont {Do}, \citenamefont
  {Kwon},\ and\ \citenamefont {Choi}}]{Glamazda2017_Raman_hysteresis}%
  \BibitemOpen
  \bibfield  {author} {\bibinfo {author} {\bibfnamefont {A.}~\bibnamefont
  {Glamazda}}, \bibinfo {author} {\bibfnamefont {P.}~\bibnamefont {Lemmens}},
  \bibinfo {author} {\bibfnamefont {S.-H.}\ \bibnamefont {Do}}, \bibinfo
  {author} {\bibfnamefont {Y.~S.}\ \bibnamefont {Kwon}}, \ and\ \bibinfo
  {author} {\bibfnamefont {K.-Y.}\ \bibnamefont {Choi}},\ }\href {\doibase
  10.1103/PhysRevB.95.174429} {\bibfield  {journal} {\bibinfo  {journal} {Phys.
  Rev. B}\ }\textbf {\bibinfo {volume} {95}},\ \bibinfo {pages} {174429}
  (\bibinfo {year} {2017})}\BibitemShut {NoStop}%
\bibitem [{\citenamefont {Johnson}\ \emph {et~al.}(2015)\citenamefont
  {Johnson}, \citenamefont {Williams}, \citenamefont {Haghighirad},
  \citenamefont {Singleton}, \citenamefont {Zapf}, \citenamefont {Manuel},
  \citenamefont {Mazin}, \citenamefont {Li}, \citenamefont {Jeschke},
  \citenamefont {Valent\'{\i}},\ and\ \citenamefont
  {Coldea}}]{Johnson2015_zigzag}%
  \BibitemOpen
  \bibfield  {author} {\bibinfo {author} {\bibfnamefont {R.~D.}\ \bibnamefont
  {Johnson}}, \bibinfo {author} {\bibfnamefont {S.~C.}\ \bibnamefont
  {Williams}}, \bibinfo {author} {\bibfnamefont {A.~A.}\ \bibnamefont
  {Haghighirad}}, \bibinfo {author} {\bibfnamefont {J.}~\bibnamefont
  {Singleton}}, \bibinfo {author} {\bibfnamefont {V.}~\bibnamefont {Zapf}},
  \bibinfo {author} {\bibfnamefont {P.}~\bibnamefont {Manuel}}, \bibinfo
  {author} {\bibfnamefont {I.~I.}\ \bibnamefont {Mazin}}, \bibinfo {author}
  {\bibfnamefont {Y.}~\bibnamefont {Li}}, \bibinfo {author} {\bibfnamefont
  {H.~O.}\ \bibnamefont {Jeschke}}, \bibinfo {author} {\bibfnamefont
  {R.}~\bibnamefont {Valent\'{\i}}}, \ and\ \bibinfo {author} {\bibfnamefont
  {R.}~\bibnamefont {Coldea}},\ }\href {\doibase 10.1103/PhysRevB.92.235119}
  {\bibfield  {journal} {\bibinfo  {journal} {Phys. Rev. B}\ }\textbf {\bibinfo
  {volume} {92}},\ \bibinfo {pages} {235119} (\bibinfo {year}
  {2015})}\BibitemShut {NoStop}%
\bibitem [{\citenamefont {Kubota}\ \emph {et~al.}(2015)\citenamefont {Kubota},
  \citenamefont {Tanaka}, \citenamefont {Ono}, \citenamefont {Narumi},\ and\
  \citenamefont {Kindo}}]{Kubota2015_sus_hysteresis}%
  \BibitemOpen
  \bibfield  {author} {\bibinfo {author} {\bibfnamefont {Y.}~\bibnamefont
  {Kubota}}, \bibinfo {author} {\bibfnamefont {H.}~\bibnamefont {Tanaka}},
  \bibinfo {author} {\bibfnamefont {T.}~\bibnamefont {Ono}}, \bibinfo {author}
  {\bibfnamefont {Y.}~\bibnamefont {Narumi}}, \ and\ \bibinfo {author}
  {\bibfnamefont {K.}~\bibnamefont {Kindo}},\ }\href {\doibase
  10.1103/PhysRevB.91.094422} {\bibfield  {journal} {\bibinfo  {journal} {Phys.
  Rev. B}\ }\textbf {\bibinfo {volume} {91}},\ \bibinfo {pages} {094422}
  (\bibinfo {year} {2015})}\BibitemShut {NoStop}%
\bibitem [{\citenamefont {Park}\ \emph {et~al.}()\citenamefont {Park},
  \citenamefont {Do}, \citenamefont {Choi}, \citenamefont {Jang}, \citenamefont
  {Jang}, \citenamefont {Schefer}, \citenamefont {Wu}, \citenamefont {Gardner},
  \citenamefont {Park}, \citenamefont {Park},\ and\ \citenamefont
  {Ji}}]{Park2016_x_ray_hysteresis}%
  \BibitemOpen
  \bibfield  {author} {\bibinfo {author} {\bibfnamefont {S.~Y.}\ \bibnamefont
  {Park}}, \bibinfo {author} {\bibfnamefont {S.~H.}\ \bibnamefont {Do}},
  \bibinfo {author} {\bibfnamefont {K.~Y.}\ \bibnamefont {Choi}}, \bibinfo
  {author} {\bibfnamefont {D.}~\bibnamefont {Jang}}, \bibinfo {author}
  {\bibfnamefont {T.~H.}\ \bibnamefont {Jang}}, \bibinfo {author}
  {\bibfnamefont {J.}~\bibnamefont {Schefer}}, \bibinfo {author} {\bibfnamefont
  {C.~M.}\ \bibnamefont {Wu}}, \bibinfo {author} {\bibfnamefont {J.~S.}\
  \bibnamefont {Gardner}}, \bibinfo {author} {\bibfnamefont {J.~M.~S.}\
  \bibnamefont {Park}}, \bibinfo {author} {\bibfnamefont {J.~H.}\ \bibnamefont
  {Park}}, \ and\ \bibinfo {author} {\bibfnamefont {S.}~\bibnamefont {Ji}},\
  }\href {http://arxiv.org/abs/1609.05690} {\ }\Eprint
  {http://arxiv.org/abs/1609.05690} {arXiv:1609.05690} \BibitemShut {NoStop}%
\bibitem [{\citenamefont {Wang}\ \emph {et~al.}(2018)\citenamefont {Wang},
  \citenamefont {Guo}, \citenamefont {Tafti}, \citenamefont {Hegg},
  \citenamefont {Sen}, \citenamefont {Sidorov}, \citenamefont {Wang},
  \citenamefont {Cai}, \citenamefont {Yi}, \citenamefont {Zhou}, \citenamefont
  {Wang}, \citenamefont {Zhang}, \citenamefont {Yang}, \citenamefont {Li},
  \citenamefont {Li}, \citenamefont {Li}, \citenamefont {Liu}, \citenamefont
  {Shi}, \citenamefont {Ku}, \citenamefont {Wu}, \citenamefont {Cava},\ and\
  \citenamefont {Sun}}]{Wang2017_pressure}%
  \BibitemOpen
  \bibfield  {author} {\bibinfo {author} {\bibfnamefont {Z.}~\bibnamefont
  {Wang}}, \bibinfo {author} {\bibfnamefont {J.}~\bibnamefont {Guo}}, \bibinfo
  {author} {\bibfnamefont {F.~F.}\ \bibnamefont {Tafti}}, \bibinfo {author}
  {\bibfnamefont {A.}~\bibnamefont {Hegg}}, \bibinfo {author} {\bibfnamefont
  {S.}~\bibnamefont {Sen}}, \bibinfo {author} {\bibfnamefont {V.~A.}\
  \bibnamefont {Sidorov}}, \bibinfo {author} {\bibfnamefont {L.}~\bibnamefont
  {Wang}}, \bibinfo {author} {\bibfnamefont {S.}~\bibnamefont {Cai}}, \bibinfo
  {author} {\bibfnamefont {W.}~\bibnamefont {Yi}}, \bibinfo {author}
  {\bibfnamefont {Y.}~\bibnamefont {Zhou}}, \bibinfo {author} {\bibfnamefont
  {H.}~\bibnamefont {Wang}}, \bibinfo {author} {\bibfnamefont {S.}~\bibnamefont
  {Zhang}}, \bibinfo {author} {\bibfnamefont {K.}~\bibnamefont {Yang}},
  \bibinfo {author} {\bibfnamefont {A.}~\bibnamefont {Li}}, \bibinfo {author}
  {\bibfnamefont {X.}~\bibnamefont {Li}}, \bibinfo {author} {\bibfnamefont
  {Y.}~\bibnamefont {Li}}, \bibinfo {author} {\bibfnamefont {J.}~\bibnamefont
  {Liu}}, \bibinfo {author} {\bibfnamefont {Y.}~\bibnamefont {Shi}}, \bibinfo
  {author} {\bibfnamefont {W.}~\bibnamefont {Ku}}, \bibinfo {author}
  {\bibfnamefont {Q.}~\bibnamefont {Wu}}, \bibinfo {author} {\bibfnamefont
  {R.~J.}\ \bibnamefont {Cava}}, \ and\ \bibinfo {author} {\bibfnamefont
  {L.}~\bibnamefont {Sun}},\ }\href {\doibase 10.1103/PhysRevB.97.245149}
  {\bibfield  {journal} {\bibinfo  {journal} {Phys. Rev. B}\ }\textbf {\bibinfo
  {volume} {97}},\ \bibinfo {pages} {245149} (\bibinfo {year}
  {2018})}\BibitemShut {NoStop}%
\bibitem [{\citenamefont {Baek}\ \emph {et~al.}(2017)\citenamefont {Baek},
  \citenamefont {Do}, \citenamefont {Choi}, \citenamefont {Kwon}, \citenamefont
  {Wolter}, \citenamefont {Nishimoto}, \citenamefont {van~den Brink},\ and\
  \citenamefont {B\"uchner}}]{Baek2017_NMR}%
  \BibitemOpen
  \bibfield  {author} {\bibinfo {author} {\bibfnamefont {S.-H.}\ \bibnamefont
  {Baek}}, \bibinfo {author} {\bibfnamefont {S.-H.}\ \bibnamefont {Do}},
  \bibinfo {author} {\bibfnamefont {K.-Y.}\ \bibnamefont {Choi}}, \bibinfo
  {author} {\bibfnamefont {Y.~S.}\ \bibnamefont {Kwon}}, \bibinfo {author}
  {\bibfnamefont {A.~U.~B.}\ \bibnamefont {Wolter}}, \bibinfo {author}
  {\bibfnamefont {S.}~\bibnamefont {Nishimoto}}, \bibinfo {author}
  {\bibfnamefont {J.}~\bibnamefont {van~den Brink}}, \ and\ \bibinfo {author}
  {\bibfnamefont {B.}~\bibnamefont {B\"uchner}},\ }\href {\doibase
  10.1103/PhysRevLett.119.037201} {\bibfield  {journal} {\bibinfo  {journal}
  {Phys. Rev. Lett.}\ }\textbf {\bibinfo {volume} {119}},\ \bibinfo {pages}
  {037201} (\bibinfo {year} {2017})}\BibitemShut {NoStop}%
\bibitem [{\citenamefont {Banerjee}\ \emph {et~al.}(2017)\citenamefont
  {Banerjee}, \citenamefont {Yan}, \citenamefont {Knolle}, \citenamefont
  {Bridges}, \citenamefont {Stone}, \citenamefont {Lumsden}, \citenamefont
  {Mandrus}, \citenamefont {Tennant}, \citenamefont {Moessner},\ and\
  \citenamefont {Nagler}}]{Banerjee2017}%
  \BibitemOpen
  \bibfield  {author} {\bibinfo {author} {\bibfnamefont {A.}~\bibnamefont
  {Banerjee}}, \bibinfo {author} {\bibfnamefont {J.}~\bibnamefont {Yan}},
  \bibinfo {author} {\bibfnamefont {J.}~\bibnamefont {Knolle}}, \bibinfo
  {author} {\bibfnamefont {C.~A.}\ \bibnamefont {Bridges}}, \bibinfo {author}
  {\bibfnamefont {M.~B.}\ \bibnamefont {Stone}}, \bibinfo {author}
  {\bibfnamefont {M.~D.}\ \bibnamefont {Lumsden}}, \bibinfo {author}
  {\bibfnamefont {D.~G.}\ \bibnamefont {Mandrus}}, \bibinfo {author}
  {\bibfnamefont {D.~A.}\ \bibnamefont {Tennant}}, \bibinfo {author}
  {\bibfnamefont {R.}~\bibnamefont {Moessner}}, \ and\ \bibinfo {author}
  {\bibfnamefont {S.~E.}\ \bibnamefont {Nagler}},\ }\href {\doibase
  10.1126/science.aah6015} {\bibfield  {journal} {\bibinfo  {journal}
  {Science}\ }\textbf {\bibinfo {volume} {356}},\ \bibinfo {pages} {1055}
  (\bibinfo {year} {2017})}\BibitemShut {NoStop}%
\bibitem [{\citenamefont {Majumder}\ \emph {et~al.}(2015)\citenamefont
  {Majumder}, \citenamefont {Schmidt}, \citenamefont {Rosner}, \citenamefont
  {Tsirlin}, \citenamefont {Yasuoka},\ and\ \citenamefont
  {Baenitz}}]{Majumder2015}%
  \BibitemOpen
  \bibfield  {author} {\bibinfo {author} {\bibfnamefont {M.}~\bibnamefont
  {Majumder}}, \bibinfo {author} {\bibfnamefont {M.}~\bibnamefont {Schmidt}},
  \bibinfo {author} {\bibfnamefont {H.}~\bibnamefont {Rosner}}, \bibinfo
  {author} {\bibfnamefont {A.~A.}\ \bibnamefont {Tsirlin}}, \bibinfo {author}
  {\bibfnamefont {H.}~\bibnamefont {Yasuoka}}, \ and\ \bibinfo {author}
  {\bibfnamefont {M.}~\bibnamefont {Baenitz}},\ }\href {\doibase
  10.1103/PhysRevB.91.180401} {\bibfield  {journal} {\bibinfo  {journal} {Phys.
  Rev. B}\ }\textbf {\bibinfo {volume} {91}},\ \bibinfo {pages} {180401}
  (\bibinfo {year} {2015})}\BibitemShut {NoStop}%
\bibitem [{\citenamefont {Kasahara}\ \emph {et~al.}(2018)\citenamefont
  {Kasahara}, \citenamefont {Ohnishi}, \citenamefont {Mizukami}, \citenamefont
  {Tanaka}, \citenamefont {Ma}, \citenamefont {Sugii}, \citenamefont {Kurita},
  \citenamefont {Tanaka}, \citenamefont {Nasu}, \citenamefont {Motome},
  \citenamefont {Shibauchi},\ and\ \citenamefont {Matsuda}}]{Kasahara:2018aa}%
  \BibitemOpen
  \bibfield  {author} {\bibinfo {author} {\bibfnamefont {Y.}~\bibnamefont
  {Kasahara}}, \bibinfo {author} {\bibfnamefont {T.}~\bibnamefont {Ohnishi}},
  \bibinfo {author} {\bibfnamefont {Y.}~\bibnamefont {Mizukami}}, \bibinfo
  {author} {\bibfnamefont {O.}~\bibnamefont {Tanaka}}, \bibinfo {author}
  {\bibfnamefont {S.}~\bibnamefont {Ma}}, \bibinfo {author} {\bibfnamefont
  {K.}~\bibnamefont {Sugii}}, \bibinfo {author} {\bibfnamefont
  {N.}~\bibnamefont {Kurita}}, \bibinfo {author} {\bibfnamefont
  {H.}~\bibnamefont {Tanaka}}, \bibinfo {author} {\bibfnamefont
  {J.}~\bibnamefont {Nasu}}, \bibinfo {author} {\bibfnamefont {Y.}~\bibnamefont
  {Motome}}, \bibinfo {author} {\bibfnamefont {T.}~\bibnamefont {Shibauchi}}, \
  and\ \bibinfo {author} {\bibfnamefont {Y.}~\bibnamefont {Matsuda}},\ }\href
  {\doibase 10.1038/s41586-018-0274-0} {\bibfield  {journal} {\bibinfo
  {journal} {Nature}\ }\textbf {\bibinfo {volume} {559}},\ \bibinfo {pages}
  {227} (\bibinfo {year} {2018})}\BibitemShut {NoStop}%
\bibitem [{\citenamefont {Norman}(2016)}]{Norman2106_review}%
  \BibitemOpen
  \bibfield  {author} {\bibinfo {author} {\bibfnamefont {M.~R.}\ \bibnamefont
  {Norman}},\ }\href {\doibase 10.1103/RevModPhys.88.041002} {\bibfield
  {journal} {\bibinfo  {journal} {Rev. Mod. Phys.}\ }\textbf {\bibinfo {volume}
  {88}},\ \bibinfo {pages} {041002} (\bibinfo {year} {2016})}\BibitemShut
  {NoStop}%
\bibitem [{\citenamefont {Zhou}\ \emph {et~al.}(2017)\citenamefont {Zhou},
  \citenamefont {Kanoda},\ and\ \citenamefont {Ng}}]{Zhou2017_review}%
  \BibitemOpen
  \bibfield  {author} {\bibinfo {author} {\bibfnamefont {Y.}~\bibnamefont
  {Zhou}}, \bibinfo {author} {\bibfnamefont {K.}~\bibnamefont {Kanoda}}, \ and\
  \bibinfo {author} {\bibfnamefont {T.-K.}\ \bibnamefont {Ng}},\ }\href
  {\doibase 10.1103/RevModPhys.89.025003} {\bibfield  {journal} {\bibinfo
  {journal} {Rev. Mod. Phys.}\ }\textbf {\bibinfo {volume} {89}},\ \bibinfo
  {pages} {025003} (\bibinfo {year} {2017})}\BibitemShut {NoStop}%
\bibitem [{\citenamefont {Cui}\ \emph {et~al.}(2017)\citenamefont {Cui},
  \citenamefont {Zheng}, \citenamefont {Ran}, \citenamefont {Wen},
  \citenamefont {Liu}, \citenamefont {Liu}, \citenamefont {Guo},\ and\
  \citenamefont {Yu}}]{Cui2017_Press_NMR}%
  \BibitemOpen
  \bibfield  {author} {\bibinfo {author} {\bibfnamefont {Y.}~\bibnamefont
  {Cui}}, \bibinfo {author} {\bibfnamefont {J.}~\bibnamefont {Zheng}}, \bibinfo
  {author} {\bibfnamefont {K.}~\bibnamefont {Ran}}, \bibinfo {author}
  {\bibfnamefont {J.}~\bibnamefont {Wen}}, \bibinfo {author} {\bibfnamefont
  {Z.-X.}\ \bibnamefont {Liu}}, \bibinfo {author} {\bibfnamefont
  {B.}~\bibnamefont {Liu}}, \bibinfo {author} {\bibfnamefont {W.}~\bibnamefont
  {Guo}}, \ and\ \bibinfo {author} {\bibfnamefont {W.}~\bibnamefont {Yu}},\
  }\href {\doibase 10.1103/PhysRevB.96.205147} {\bibfield  {journal} {\bibinfo
  {journal} {Phys. Rev. B}\ }\textbf {\bibinfo {volume} {96}},\ \bibinfo
  {pages} {205147} (\bibinfo {year} {2017})}\BibitemShut {NoStop}%
\bibitem [{\citenamefont {He}\ \emph {et~al.}(2018)\citenamefont {He},
  \citenamefont {Wang}, \citenamefont {Wang}, \citenamefont {Hardy},
  \citenamefont {Wolf}, \citenamefont {Adelmann}, \citenamefont {Br{\"u}ckel},
  \citenamefont {Su},\ and\ \citenamefont {Meingast}}]{He_2018}%
  \BibitemOpen
  \bibfield  {author} {\bibinfo {author} {\bibfnamefont {M.}~\bibnamefont
  {He}}, \bibinfo {author} {\bibfnamefont {X.}~\bibnamefont {Wang}}, \bibinfo
  {author} {\bibfnamefont {L.}~\bibnamefont {Wang}}, \bibinfo {author}
  {\bibfnamefont {F.}~\bibnamefont {Hardy}}, \bibinfo {author} {\bibfnamefont
  {T.}~\bibnamefont {Wolf}}, \bibinfo {author} {\bibfnamefont {P.}~\bibnamefont
  {Adelmann}}, \bibinfo {author} {\bibfnamefont {T.}~\bibnamefont
  {Br{\"u}ckel}}, \bibinfo {author} {\bibfnamefont {Y.}~\bibnamefont {Su}}, \
  and\ \bibinfo {author} {\bibfnamefont {C.}~\bibnamefont {Meingast}},\ }\href
  {\doibase 10.1088/1361-648x/aada1e} {\bibfield  {journal} {\bibinfo
  {journal} {J. Phys.: Condens. Matter}\ }\textbf {\bibinfo {volume} {30}},\
  \bibinfo {pages} {385702} (\bibinfo {year} {2018})}\BibitemShut {NoStop}%
\bibitem [{\citenamefont {Banerjee}\ \emph {et~al.}(2018)\citenamefont
  {Banerjee}, \citenamefont {Lampen-Kelley}, \citenamefont {Knolle},
  \citenamefont {Balz}, \citenamefont {Aczel}, \citenamefont {Winn},
  \citenamefont {Liu}, \citenamefont {Pajerowski}, \citenamefont {Yan},
  \citenamefont {Bridges}, \citenamefont {Savici}, \citenamefont {Chakoumakos},
  \citenamefont {Lumsden}, \citenamefont {Tennant}, \citenamefont {Moessner},
  \citenamefont {Mandrus},\ and\ \citenamefont {Nagler}}]{Banerjee2017_field}%
  \BibitemOpen
  \bibfield  {author} {\bibinfo {author} {\bibfnamefont {A.}~\bibnamefont
  {Banerjee}}, \bibinfo {author} {\bibfnamefont {P.}~\bibnamefont
  {Lampen-Kelley}}, \bibinfo {author} {\bibfnamefont {J.}~\bibnamefont
  {Knolle}}, \bibinfo {author} {\bibfnamefont {C.}~\bibnamefont {Balz}},
  \bibinfo {author} {\bibfnamefont {A.~A.}\ \bibnamefont {Aczel}}, \bibinfo
  {author} {\bibfnamefont {B.}~\bibnamefont {Winn}}, \bibinfo {author}
  {\bibfnamefont {Y.}~\bibnamefont {Liu}}, \bibinfo {author} {\bibfnamefont
  {D.}~\bibnamefont {Pajerowski}}, \bibinfo {author} {\bibfnamefont
  {J.}~\bibnamefont {Yan}}, \bibinfo {author} {\bibfnamefont {C.~A.}\
  \bibnamefont {Bridges}}, \bibinfo {author} {\bibfnamefont {A.~T.}\
  \bibnamefont {Savici}}, \bibinfo {author} {\bibfnamefont {B.~C.}\
  \bibnamefont {Chakoumakos}}, \bibinfo {author} {\bibfnamefont {M.~D.}\
  \bibnamefont {Lumsden}}, \bibinfo {author} {\bibfnamefont {D.~A.}\
  \bibnamefont {Tennant}}, \bibinfo {author} {\bibfnamefont {R.}~\bibnamefont
  {Moessner}}, \bibinfo {author} {\bibfnamefont {D.~G.}\ \bibnamefont
  {Mandrus}}, \ and\ \bibinfo {author} {\bibfnamefont {S.~E.}\ \bibnamefont
  {Nagler}},\ }\href {\doibase 10.1038/s41535-018-0079-2} {\bibfield  {journal}
  {\bibinfo  {journal} {npj Quantum Mater.}\ }\textbf {\bibinfo {volume} {3}},\
  \bibinfo {pages} {8} (\bibinfo {year} {2018})}\BibitemShut {NoStop}%
\bibitem [{\citenamefont {Gass}\ \emph {et~al.}(2020)\citenamefont {Gass},
  \citenamefont {C\^onsoli}, \citenamefont {Kocsis}, \citenamefont {Corredor},
  \citenamefont {Lampen-Kelley}, \citenamefont {Mandrus}, \citenamefont
  {Nagler}, \citenamefont {Janssen}, \citenamefont {Vojta}, \citenamefont
  {B\"uchner},\ and\ \citenamefont {Wolter}}]{Gass2020}%
  \BibitemOpen
  \bibfield  {author} {\bibinfo {author} {\bibfnamefont {S.}~\bibnamefont
  {Gass}}, \bibinfo {author} {\bibfnamefont {P.~M.}\ \bibnamefont {C\^onsoli}},
  \bibinfo {author} {\bibfnamefont {V.}~\bibnamefont {Kocsis}}, \bibinfo
  {author} {\bibfnamefont {L.~T.}\ \bibnamefont {Corredor}}, \bibinfo {author}
  {\bibfnamefont {P.}~\bibnamefont {Lampen-Kelley}}, \bibinfo {author}
  {\bibfnamefont {D.~G.}\ \bibnamefont {Mandrus}}, \bibinfo {author}
  {\bibfnamefont {S.~E.}\ \bibnamefont {Nagler}}, \bibinfo {author}
  {\bibfnamefont {L.}~\bibnamefont {Janssen}}, \bibinfo {author} {\bibfnamefont
  {M.}~\bibnamefont {Vojta}}, \bibinfo {author} {\bibfnamefont
  {B.}~\bibnamefont {B\"uchner}}, \ and\ \bibinfo {author} {\bibfnamefont
  {A.~U.~B.}\ \bibnamefont {Wolter}},\ }\href {\doibase
  10.1103/PhysRevB.101.245158} {\bibfield  {journal} {\bibinfo  {journal}
  {Phys. Rev. B}\ }\textbf {\bibinfo {volume} {101}},\ \bibinfo {pages}
  {245158} (\bibinfo {year} {2020})}\BibitemShut {NoStop}%
\bibitem [{\citenamefont {Balz}\ \emph {et~al.}(2019)\citenamefont {Balz},
  \citenamefont {Lampen-Kelley}, \citenamefont {Banerjee}, \citenamefont {Yan},
  \citenamefont {Lu}, \citenamefont {Hu}, \citenamefont {Yadav}, \citenamefont
  {Takano}, \citenamefont {Liu}, \citenamefont {Tennant}, \citenamefont
  {Lumsden}, \citenamefont {Mandrus},\ and\ \citenamefont {Nagler}}]{Balz2019}%
  \BibitemOpen
  \bibfield  {author} {\bibinfo {author} {\bibfnamefont {C.}~\bibnamefont
  {Balz}}, \bibinfo {author} {\bibfnamefont {P.}~\bibnamefont {Lampen-Kelley}},
  \bibinfo {author} {\bibfnamefont {A.}~\bibnamefont {Banerjee}}, \bibinfo
  {author} {\bibfnamefont {J.}~\bibnamefont {Yan}}, \bibinfo {author}
  {\bibfnamefont {Z.}~\bibnamefont {Lu}}, \bibinfo {author} {\bibfnamefont
  {X.}~\bibnamefont {Hu}}, \bibinfo {author} {\bibfnamefont {S.~M.}\
  \bibnamefont {Yadav}}, \bibinfo {author} {\bibfnamefont {Y.}~\bibnamefont
  {Takano}}, \bibinfo {author} {\bibfnamefont {Y.}~\bibnamefont {Liu}},
  \bibinfo {author} {\bibfnamefont {D.~A.}\ \bibnamefont {Tennant}}, \bibinfo
  {author} {\bibfnamefont {M.~D.}\ \bibnamefont {Lumsden}}, \bibinfo {author}
  {\bibfnamefont {D.}~\bibnamefont {Mandrus}}, \ and\ \bibinfo {author}
  {\bibfnamefont {S.~E.}\ \bibnamefont {Nagler}},\ }\href {\doibase
  10.1103/PhysRevB.100.060405} {\bibfield  {journal} {\bibinfo  {journal}
  {Phys. Rev. B}\ }\textbf {\bibinfo {volume} {100}},\ \bibinfo {pages}
  {060405} (\bibinfo {year} {2019})}\BibitemShut {NoStop}%
\bibitem [{\citenamefont {Lampen-Kelley}\ \emph {et~al.}()\citenamefont
  {Lampen-Kelley}, \citenamefont {Janssen}, \citenamefont {Andrade},
  \citenamefont {Rachel}, \citenamefont {Yan}, \citenamefont {Balz},
  \citenamefont {Mandrus}, \citenamefont {Nagler},\ and\ \citenamefont
  {Vojta}}]{Lampen-Kelley2018}%
  \BibitemOpen
  \bibfield  {author} {\bibinfo {author} {\bibfnamefont {P.}~\bibnamefont
  {Lampen-Kelley}}, \bibinfo {author} {\bibfnamefont {L.}~\bibnamefont
  {Janssen}}, \bibinfo {author} {\bibfnamefont {E.~C.}\ \bibnamefont
  {Andrade}}, \bibinfo {author} {\bibfnamefont {S.}~\bibnamefont {Rachel}},
  \bibinfo {author} {\bibfnamefont {J.~Q.}\ \bibnamefont {Yan}}, \bibinfo
  {author} {\bibfnamefont {C.}~\bibnamefont {Balz}}, \bibinfo {author}
  {\bibfnamefont {D.~G.}\ \bibnamefont {Mandrus}}, \bibinfo {author}
  {\bibfnamefont {S.~E.}\ \bibnamefont {Nagler}}, \ and\ \bibinfo {author}
  {\bibfnamefont {M.}~\bibnamefont {Vojta}},\ }\href
  {http://arxiv.org/abs/1807.06192} {\ }\Eprint
  {http://arxiv.org/abs/1807.06192} {arXiv:1807.06192} \BibitemShut {NoStop}%
\bibitem [{\citenamefont {Ran}\ \emph {et~al.}(2017)\citenamefont {Ran},
  \citenamefont {Wang}, \citenamefont {Wang}, \citenamefont {Dong},
  \citenamefont {Ren}, \citenamefont {Bao}, \citenamefont {Li}, \citenamefont
  {Ma}, \citenamefont {Gan}, \citenamefont {Zhang}, \citenamefont {Park},
  \citenamefont {Deng}, \citenamefont {Danilkin}, \citenamefont {Yu},
  \citenamefont {Li},\ and\ \citenamefont {Wen}}]{Ran2017}%
  \BibitemOpen
  \bibfield  {author} {\bibinfo {author} {\bibfnamefont {K.}~\bibnamefont
  {Ran}}, \bibinfo {author} {\bibfnamefont {J.}~\bibnamefont {Wang}}, \bibinfo
  {author} {\bibfnamefont {W.}~\bibnamefont {Wang}}, \bibinfo {author}
  {\bibfnamefont {Z.-Y.}\ \bibnamefont {Dong}}, \bibinfo {author}
  {\bibfnamefont {X.}~\bibnamefont {Ren}}, \bibinfo {author} {\bibfnamefont
  {S.}~\bibnamefont {Bao}}, \bibinfo {author} {\bibfnamefont {S.}~\bibnamefont
  {Li}}, \bibinfo {author} {\bibfnamefont {Z.}~\bibnamefont {Ma}}, \bibinfo
  {author} {\bibfnamefont {Y.}~\bibnamefont {Gan}}, \bibinfo {author}
  {\bibfnamefont {Y.}~\bibnamefont {Zhang}}, \bibinfo {author} {\bibfnamefont
  {J.~T.}\ \bibnamefont {Park}}, \bibinfo {author} {\bibfnamefont
  {G.}~\bibnamefont {Deng}}, \bibinfo {author} {\bibfnamefont {S.}~\bibnamefont
  {Danilkin}}, \bibinfo {author} {\bibfnamefont {S.-L.}\ \bibnamefont {Yu}},
  \bibinfo {author} {\bibfnamefont {J.-X.}\ \bibnamefont {Li}}, \ and\ \bibinfo
  {author} {\bibfnamefont {J.}~\bibnamefont {Wen}},\ }\href {\doibase
  10.1103/PhysRevLett.118.107203} {\bibfield  {journal} {\bibinfo  {journal}
  {Phys. Rev. Lett.}\ }\textbf {\bibinfo {volume} {118}},\ \bibinfo {pages}
  {107203} (\bibinfo {year} {2017})}\BibitemShut {NoStop}%
\bibitem [{\citenamefont {Do}\ \emph {et~al.}(2017)\citenamefont {Do},
  \citenamefont {Park}, \citenamefont {Yoshitake}, \citenamefont {Nasu},
  \citenamefont {Motome}, \citenamefont {Kwon}, \citenamefont {Adroja},
  \citenamefont {Voneshen}, \citenamefont {Kim}, \citenamefont {Jang},
  \citenamefont {Park}, \citenamefont {Choi},\ and\ \citenamefont
  {Ji}}]{Do:2017aaHeatcapacity}%
  \BibitemOpen
  \bibfield  {author} {\bibinfo {author} {\bibfnamefont {S.-H.}\ \bibnamefont
  {Do}}, \bibinfo {author} {\bibfnamefont {S.-Y.}\ \bibnamefont {Park}},
  \bibinfo {author} {\bibfnamefont {J.}~\bibnamefont {Yoshitake}}, \bibinfo
  {author} {\bibfnamefont {J.}~\bibnamefont {Nasu}}, \bibinfo {author}
  {\bibfnamefont {Y.}~\bibnamefont {Motome}}, \bibinfo {author} {\bibfnamefont
  {Y.~S.}\ \bibnamefont {Kwon}}, \bibinfo {author} {\bibfnamefont {D.~T.}\
  \bibnamefont {Adroja}}, \bibinfo {author} {\bibfnamefont {D.~J.}\
  \bibnamefont {Voneshen}}, \bibinfo {author} {\bibfnamefont {K.}~\bibnamefont
  {Kim}}, \bibinfo {author} {\bibfnamefont {T.~H.}\ \bibnamefont {Jang}},
  \bibinfo {author} {\bibfnamefont {J.~H.}\ \bibnamefont {Park}}, \bibinfo
  {author} {\bibfnamefont {K.-Y.}\ \bibnamefont {Choi}}, \ and\ \bibinfo
  {author} {\bibfnamefont {S.}~\bibnamefont {Ji}},\ }\href {\doibase
  10.1038/nphys4264} {\bibfield  {journal} {\bibinfo  {journal} {Nat. Phys.}\
  }\textbf {\bibinfo {volume} {13}},\ \bibinfo {pages} {1079} (\bibinfo {year}
  {2017})}\BibitemShut {NoStop}%
\bibitem [{\citenamefont {Balz}\ \emph {et~al.}()\citenamefont {Balz},
  \citenamefont {Janssen}, \citenamefont {Lampen-Kelley}, \citenamefont
  {Banerjee}, \citenamefont {Liu}, \citenamefont {Yan}, \citenamefont
  {Mandrus}, \citenamefont {Vojta},\ and\ \citenamefont {Nagler}}]{Balz2020}%
  \BibitemOpen
  \bibfield  {author} {\bibinfo {author} {\bibfnamefont {C.}~\bibnamefont
  {Balz}}, \bibinfo {author} {\bibfnamefont {L.}~\bibnamefont {Janssen}},
  \bibinfo {author} {\bibfnamefont {P.}~\bibnamefont {Lampen-Kelley}}, \bibinfo
  {author} {\bibfnamefont {A.}~\bibnamefont {Banerjee}}, \bibinfo {author}
  {\bibfnamefont {Y.}~\bibnamefont {Liu}}, \bibinfo {author} {\bibfnamefont
  {J.}~\bibnamefont {Yan}}, \bibinfo {author} {\bibfnamefont {D.}~\bibnamefont
  {Mandrus}}, \bibinfo {author} {\bibfnamefont {M.}~\bibnamefont {Vojta}}, \
  and\ \bibinfo {author} {\bibfnamefont {S.~E.}\ \bibnamefont {Nagler}},\
  }\href {http://arxiv.org/abs/2012.15258} {\ }\Eprint
  {http://arxiv.org/abs/2012.15258} {arXiv:2012.15258} \BibitemShut {NoStop}%
\bibitem [{\citenamefont {Kim}\ and\ \citenamefont
  {Kee}(2016)}]{Kim2016_weak_bonding}%
  \BibitemOpen
  \bibfield  {author} {\bibinfo {author} {\bibfnamefont {H.-S.}\ \bibnamefont
  {Kim}}\ and\ \bibinfo {author} {\bibfnamefont {H.-Y.}\ \bibnamefont {Kee}},\
  }\href {\doibase 10.1103/PhysRevB.93.155143} {\bibfield  {journal} {\bibinfo
  {journal} {Phys. Rev. B}\ }\textbf {\bibinfo {volume} {93}},\ \bibinfo
  {pages} {155143} (\bibinfo {year} {2016})}\BibitemShut {NoStop}%
\bibitem [{\citenamefont {Widmann}\ \emph {et~al.}(2019)\citenamefont
  {Widmann}, \citenamefont {Tsurkan}, \citenamefont {Prishchenko},
  \citenamefont {Mazurenko}, \citenamefont {Tsirlin},\ and\ \citenamefont
  {Loidl}}]{Widmann2019_Thermalexpansion}%
  \BibitemOpen
  \bibfield  {author} {\bibinfo {author} {\bibfnamefont {S.}~\bibnamefont
  {Widmann}}, \bibinfo {author} {\bibfnamefont {V.}~\bibnamefont {Tsurkan}},
  \bibinfo {author} {\bibfnamefont {D.~A.}\ \bibnamefont {Prishchenko}},
  \bibinfo {author} {\bibfnamefont {V.~G.}\ \bibnamefont {Mazurenko}}, \bibinfo
  {author} {\bibfnamefont {A.~A.}\ \bibnamefont {Tsirlin}}, \ and\ \bibinfo
  {author} {\bibfnamefont {A.}~\bibnamefont {Loidl}},\ }\href {\doibase
  10.1103/PhysRevB.99.094415} {\bibfield  {journal} {\bibinfo  {journal} {Phys.
  Rev. B}\ }\textbf {\bibinfo {volume} {99}},\ \bibinfo {pages} {094415}
  (\bibinfo {year} {2019})}\BibitemShut {NoStop}%
\bibitem [{\citenamefont {Reschke}\ \emph {et~al.}(2018)\citenamefont
  {Reschke}, \citenamefont {Mayr}, \citenamefont {Widmann}, \citenamefont {von
  Nidda}, \citenamefont {Tsurkan}, \citenamefont {Eremin}, \citenamefont {Do},
  \citenamefont {Choi}, \citenamefont {Wang},\ and\ \citenamefont
  {Loidl}}]{Reschke_2018_thermalexpansion}%
  \BibitemOpen
  \bibfield  {author} {\bibinfo {author} {\bibfnamefont {S.}~\bibnamefont
  {Reschke}}, \bibinfo {author} {\bibfnamefont {F.}~\bibnamefont {Mayr}},
  \bibinfo {author} {\bibfnamefont {S.}~\bibnamefont {Widmann}}, \bibinfo
  {author} {\bibfnamefont {H.-A.~K.}\ \bibnamefont {von Nidda}}, \bibinfo
  {author} {\bibfnamefont {V.}~\bibnamefont {Tsurkan}}, \bibinfo {author}
  {\bibfnamefont {M.~V.}\ \bibnamefont {Eremin}}, \bibinfo {author}
  {\bibfnamefont {S.-H.}\ \bibnamefont {Do}}, \bibinfo {author} {\bibfnamefont
  {K.-Y.}\ \bibnamefont {Choi}}, \bibinfo {author} {\bibfnamefont
  {Z.}~\bibnamefont {Wang}}, \ and\ \bibinfo {author} {\bibfnamefont
  {A.}~\bibnamefont {Loidl}},\ }\href {\doibase 10.1088/1361-648x/aae805}
  {\bibfield  {journal} {\bibinfo  {journal} {J. Phys.: Condens. Matter}\
  }\textbf {\bibinfo {volume} {30}},\ \bibinfo {pages} {475604} (\bibinfo
  {year} {2018})}\BibitemShut {NoStop}%
\bibitem [{\citenamefont {Zheng}\ \emph {et~al.}(2018)\citenamefont {Zheng},
  \citenamefont {Cui}, \citenamefont {Li}, \citenamefont {Ran}, \citenamefont
  {Wen},\ and\ \citenamefont {Yu}}]{Zheng:2018aa}%
  \BibitemOpen
  \bibfield  {author} {\bibinfo {author} {\bibfnamefont {J.}~\bibnamefont
  {Zheng}}, \bibinfo {author} {\bibfnamefont {Y.}~\bibnamefont {Cui}}, \bibinfo
  {author} {\bibfnamefont {T.}~\bibnamefont {Li}}, \bibinfo {author}
  {\bibfnamefont {K.}~\bibnamefont {Ran}}, \bibinfo {author} {\bibfnamefont
  {J.}~\bibnamefont {Wen}}, \ and\ \bibinfo {author} {\bibfnamefont
  {W.}~\bibnamefont {Yu}},\ }\href {\doibase 10.1007/s11433-017-9166-1}
  {\bibfield  {journal} {\bibinfo  {journal} {Sci. China-Phys. Mech. Astron.}\
  }\textbf {\bibinfo {volume} {61}},\ \bibinfo {pages} {057021} (\bibinfo
  {year} {2018})}\BibitemShut {NoStop}%
\bibitem [{\citenamefont {Stroganov}\ and\ \citenamefont
  {Ovchinnikov}(1957)}]{Stroganov1957}%
  \BibitemOpen
  \bibfield  {author} {\bibinfo {author} {\bibfnamefont {E.~V.}\ \bibnamefont
  {Stroganov}}\ and\ \bibinfo {author} {\bibfnamefont {K.~V.}\ \bibnamefont
  {Ovchinnikov}},\ }\href@noop {} {\bibfield  {journal} {\bibinfo  {journal}
  {Ser. Fiz. Khim.}\ }\textbf {\bibinfo {volume} {12}},\ \bibinfo {pages} {152}
  (\bibinfo {year} {1957})}\BibitemShut {NoStop}%
\bibitem [{\citenamefont {Aoyama}\ \emph {et~al.}(2017)\citenamefont {Aoyama},
  \citenamefont {Hasegawa}, \citenamefont {Kimura}, \citenamefont {Kimura},\
  and\ \citenamefont {Ohgushi}}]{Takuya2017}%
  \BibitemOpen
  \bibfield  {author} {\bibinfo {author} {\bibfnamefont {T.}~\bibnamefont
  {Aoyama}}, \bibinfo {author} {\bibfnamefont {Y.}~\bibnamefont {Hasegawa}},
  \bibinfo {author} {\bibfnamefont {S.}~\bibnamefont {Kimura}}, \bibinfo
  {author} {\bibfnamefont {T.}~\bibnamefont {Kimura}}, \ and\ \bibinfo {author}
  {\bibfnamefont {K.}~\bibnamefont {Ohgushi}},\ }\href {\doibase
  10.1103/PhysRevB.95.245104} {\bibfield  {journal} {\bibinfo  {journal} {Phys.
  Rev. B}\ }\textbf {\bibinfo {volume} {95}},\ \bibinfo {pages} {245104}
  (\bibinfo {year} {2017})}\BibitemShut {NoStop}%
\bibitem [{\citenamefont {Courtens}(1984)}]{Courtens1984}%
  \BibitemOpen
  \bibfield  {author} {\bibinfo {author} {\bibfnamefont {E.}~\bibnamefont
  {Courtens}},\ }\href {\doibase 10.1103/PhysRevLett.52.69} {\bibfield
  {journal} {\bibinfo  {journal} {Phys. Rev. Lett.}\ }\textbf {\bibinfo
  {volume} {52}},\ \bibinfo {pages} {69} (\bibinfo {year} {1984})}\BibitemShut
  {NoStop}%
\bibitem [{\citenamefont {Bhowmik}\ and\ \citenamefont
  {Ranganathan}(2002)}]{BHOWMIK2002101}%
  \BibitemOpen
  \bibfield  {author} {\bibinfo {author} {\bibfnamefont {R.}~\bibnamefont
  {Bhowmik}}\ and\ \bibinfo {author} {\bibfnamefont {R.}~\bibnamefont
  {Ranganathan}},\ }\href {\doibase
  https://doi.org/10.1016/S0304-8853(02)00190-7} {\bibfield  {journal}
  {\bibinfo  {journal} {J. Magn. Magn. Mater.}\ }\textbf {\bibinfo {volume}
  {248}},\ \bibinfo {pages} {101} (\bibinfo {year} {2002})}\BibitemShut
  {NoStop}%
\bibitem [{\citenamefont {Bokov}\ and\ \citenamefont
  {Ye}(2006)}]{Bokov:2006aa}%
  \BibitemOpen
  \bibfield  {author} {\bibinfo {author} {\bibfnamefont {A.~A.}\ \bibnamefont
  {Bokov}}\ and\ \bibinfo {author} {\bibfnamefont {Z.~G.}\ \bibnamefont {Ye}},\
  }\href {\doibase 10.1007/s10853-005-5915-7} {\bibfield  {journal} {\bibinfo
  {journal} {J. Mater. Sci.}\ }\textbf {\bibinfo {volume} {41}},\ \bibinfo
  {pages} {31} (\bibinfo {year} {2006})}\BibitemShut {NoStop}%
\bibitem [{\citenamefont {Sears}\ \emph {et~al.}(2017)\citenamefont {Sears},
  \citenamefont {Zhao}, \citenamefont {Xu}, \citenamefont {Lynn},\ and\
  \citenamefont {Kim}}]{Sears2017Domain}%
  \BibitemOpen
  \bibfield  {author} {\bibinfo {author} {\bibfnamefont {J.~A.}\ \bibnamefont
  {Sears}}, \bibinfo {author} {\bibfnamefont {Y.}~\bibnamefont {Zhao}},
  \bibinfo {author} {\bibfnamefont {Z.}~\bibnamefont {Xu}}, \bibinfo {author}
  {\bibfnamefont {J.~W.}\ \bibnamefont {Lynn}}, \ and\ \bibinfo {author}
  {\bibfnamefont {Y.-J.}\ \bibnamefont {Kim}},\ }\href {\doibase
  10.1103/PhysRevB.95.180411} {\bibfield  {journal} {\bibinfo  {journal} {Phys.
  Rev. B}\ }\textbf {\bibinfo {volume} {95}},\ \bibinfo {pages} {180411}
  (\bibinfo {year} {2017})}\BibitemShut {NoStop}%
\end{thebibliography}

\end{document}